\documentclass[fleqn,usenatbib]{mnras}

\usepackage{newtxtext,newtxmath}

\usepackage[T1]{fontenc}
\usepackage{ae,aecompl}
\usepackage{threeparttablex}
\usepackage{footnote}

\usepackage{graphicx}	
\usepackage{subfig}
\usepackage{longtable}
\usepackage{multirow}
\usepackage{multirow,makecell}
\usepackage{array}
\usepackage{hyperref} 
\usepackage{cleveref}
\usepackage{colortbl}
\newcolumntype{P}[1]{>{\centering\arraybackslash}p{#1}}

\newcommand{\ergs}{erg\,s$^{-1}$}

\usepackage{caption}

\usepackage{tikz,xcolor,hyperref}

\definecolor{lime}{HTML}{A6CE39}
\DeclareRobustCommand{\orcidicon}{
	\begin{tikzpicture}
	\draw[lime, fill=lime] (0,0) 
	circle [radius=0.16] 
	node[white] {{\fontfamily{qag}\selectfont \tiny ID}};
	\draw[white, fill=white] (-0.0625,0.095) 
	circle [radius=0.007];
	\end{tikzpicture}
	\hspace{-2mm}
}

\foreach \x in {A, ..., Z}{\expandafter\xdef\csname orcid\x\endcsname{\noexpand\href{https://orcid.org/\csname orcidauthor\x\endcsname}
			{\noexpand\orcidicon}}
}

\title[SNR Populations of  NGC\,45, NGC\,55, NGC\,1313, NGC\,7793]{The Supernova Remnant Populations of the galaxies NGC\,45, NGC\,55, NGC\,1313, NGC\,7793:  Luminosity and Excitation Functions.}

\author[M. Kopsacheili et al.]{
M. Kopsacheili,$^{1,2}$\thanks{E-mail: mariakop@physics.uoc.gr}
A. Zezas,$^{1,2,3}$
I. Leonidaki, $^{1,2}$
P. Boumis$^{4\orcidA{}}$
\\
$^{1}$Department of Physics, University of Crete, Heraklion GR-70013, Greece \\
$^{2}$Institute of Astrophysics, FORTH, GR-71110 Heraklion, Greece\\
$^{3}$Harvard-Smithsonian Center for Astrophysics, 60 Garden Street, Cambridge, MA 02138, USA\\
$^{4}$Institute for Astronomy, Astrophysics, Space Applications
and Remote Sensing, National Observatory of Athens,
15236 Penteli, Greece}

\date{Accepted XXX. Received YYY; in original form ZZZ}

\pubyear{2021}

\begin{document}
\label{firstpage}
\pagerange{\pageref{firstpage}--\pageref{lastpage}}
\maketitle

\begin{abstract}
We present a systematic study of the Supernova Remnant (SNR) populations in the nearby galaxies NGC\,45, NGC\,55, NGC\,1313, and NGC\,7793 based on deep H$\rm{\alpha}$ and [\ion{S}{II}] imaging. We find 42 candidate and 54 possible candidate SNRs based on the [\ion{S}{II}]/H$\rm{\alpha}$>0.4 criterion, 84 of which are new identifications. We derive the  H$\rm{\alpha}$ and the joint [\ion{S}{II}]-H$\rm{\alpha}$ luminosity functions after accounting for incompleteness effects. We find that the H$\rm{\alpha}$  luminosity function of the overall sample is described with a skewed Gaussian with a mean equal to $\rm \log(L_{H\alpha}/10^{36}\, erg\,s^{-1})=0.07 $ and $\rm \sigma(\log(L_{H\alpha}/10^{36}\, erg\,s^{-1}))=0.58$. The  joint [\ion{S}{II]}-H$\rm{\alpha}$ function is parameterized  by a skewed Gaussian along the log([\ion{S}{II}]$\rm /10^{36}\, erg\,s^{-1}) = 0.88 \times \log(L_{H\alpha}/10^{36}\, erg\,s^{-1}) -  0.06$ line and a truncated Gaussian with $\rm \mu(\log(L_{[S\, II]}/10^{36})) = 0.024 $ and $\rm \sigma(\log(L_{[S\, II]}/10^{36})) = 0.14$, on its vertical direction.  
We also define the excitation function as the number density of SNRs as a function of their [\ion{S}{II}]/H$\rm{\alpha}$ ratios. This function is represented by a truncated Gaussian with a mean at -0.014. We find a sub-linear [\ion{S}{II}]-H$\rm{\alpha}$ relation indicating lower excitation for the more luminous objects. 
\end{abstract}




\section{Introduction}

Supernova remnants (SNRs) are an integral component of the interstellar medium (ISM) of a galaxy. These sources drive the evolution of the ISM through the deposition of large amounts of chemically-enriched material and kinetic energy.
 They are related to the star-forming process within a galaxy, since under appropriate conditions, their shock waves compress the ISM, triggering new-star formation. In addition, core-collapse SNRs are tracers of the on-going massive star formation  since they are the last stage in the life of massive stars ($\rm{M > 8\,M\odot}$; \citealt{1990ApJ...357...97C}).

\par In our Galaxy, 294  SNRs have been identified so far, based on the systematic study in radio bands (\citealt{2019JApA...40...36G}). Many of them have been studied in various wavelengths (e.g. \citealt{Dan2013}; \citealt{2009A&A...499..789B}; \citealt{slane2002}) yielding detailed information on their physical and morphological properties, as well as their interaction with their local ISM.  Studying SNRs in different wavelengths gives us the opportunity to examine different gas phases of SNRs, from gases with temperatures ranging from thousands to millions of degrees and through different evolutionary stages. However, the study of Galactic SNRs has some disadvantages:  optical and X-ray bands are suffering from significant extinction, leading sometimes to incomplete samples. In addition, the study of Galactic SNRs does not allow us to examine SNR populations that evolve in a wide range of ISM parameters (e.g. different density distributions, metallicities). Another important limitation of studies in Galactic SNRs is that it is very difficult to estimate their distances. The study of extragalactic SNRs remedies these limitations. They give us the opportunity to explore larger and more diverse samples that can be correlated to different ISM and galaxy properties. In addition, extragalactic sources overcome the problems of Galactic absorption and distance uncertainties, since they are considered to be at the distance of their host galaxy. Taking advantage of these benefits, many studies of SNRs in nearby galaxies have been carried out (e.g. \citealt{Long2019}; \citealt{Lee2015}; \citealt{2013MNRAS.429..189L}; \citealt{2010ApJ...725..842L}; \citealt{Blair1997}; \citealt{1997ApJS..112...49M}; \citealt{1997AAS...190.5306L}; \citealt{2000ApJ...544..780P}; \citealt{2007AJ....133.1361P}). These studies increase the number of known SNRs and thus provide significant information on their nature not only on individual objects,  but also on their overall population and their effect on their host galaxy.

\par A significant limitation of these surveys is that they do not account for the effect of incompleteness. During the detection process, there is a fraction of SNRs that we do not detect because of their faintness, the detection process, and the method we have followed for the selection and identification of SNRs. These effects make our samples incomplete.  Incompleteness is a fundamental problem that can affect significantly our results. For example, we can see features in luminosity functions which are not real and they are simply artefacts of the detection and selection process. This problem is exacerbated by the commonly used practice of SNR identification based on visual inspection of $\rm H\alpha$, [\ion{S}{II}], and [\ion{S}{II}]/$\rm H\alpha$ maps. Visual inspection, being a subjective method, does not allow the reliable quantification of the selection effects. The effect of incompleteness does not allow us to exploit the full extent of the available data. This is particularly important for studying the low luminosity end of the SNR population and hence obtaining a more complete picture of their population and their feedback to ISM. 


\par In this work, we present a systematic study of SNRs in 4 nearby galaxies: NGC\,45, NGC\,55, NGC\,1313, and NGC\,7793. We present a new way for the detection and characterization of the SNR population that bypasses the commonly used practice of visual inspection. This has the advantage of allowing the calculation of incompleteness, which in turn, enables the derivation of luminosity functions and more reliable statistics on the SNR populations. We do not restrict this analysis only to H$\rm{\alpha}$ luminosity but we also construct the joint incompleteness-corrected luminosity functions (LFs) for H$\rm{\alpha}$ and [\ion{S}{II}], the two parameters that we use for the detection. This has the additional benefit of allowing the simultaneous calculation of the SNR excitation (i.e. the [\ion{S}{II}]/H$\rm \alpha$ ratio) as a function of the H$\rm{\alpha}$ luminosity.

\par In \Cref{sample}, we describe the sample of the galaxies and their characteristics. In \Cref{Observations and Data reduction} and \Cref{Detection and Photometry} we present the observations, the data reduction, the detection of SNRs and the photometric analysis. We also describe how we calculate the incompleteness corrections. The luminosity function and the excitation as a function of the H$\rm{\alpha}$ luminosity are presented in \Cref{results} and in \Cref{discussion} we discuss our results in the context of extragalactic SNR populations. In \Cref{conclusion} we summarize our results.

\section{Sample} \label{sample}
Our sample consists of four southern galaxies: NGC\,45, NGC\,55, NGC\,1313, and NGC\,7793. Their basic properties are given in  \autoref{table:NED}. 
The main selection criteria for our sample are:
(1) They are nearby galaxies, (\rm{D<7\,Mpc})  allowing identification of individual objects; 
(2) They are face on in order to minimize the effects of internal extinction (only NGC\,55 is edge-on but its  extinction is  very low); (3) They are all spiral galaxies.
(4) They have available X-ray data from the \textit{Chandra}  X-ray observatory  (the total exposure time of the \textit{Chandra} data is presented in \autoref{table:NED}).

They also have similar metallicities, limiting metallicity effects in the characterization of SNRs.
The [O/H] metallicities given in \autoref{table:NED}, are calculated using the [\ion{N}{II}]$\rm \lambda \lambda 6548,6583$/H$\rm{\alpha}$ line ratios from the work of \citet{2008ApJS..178..247K}, which were converted to metallicities using the ($\rm 12 + log(O/H) = 8.9 + 0.59log([\ion{N}{II}]/H\alpha)$) calibration of \citet{2004MNRAS.348L..59P}. The extinction corrected H$\rm \alpha$ star formation rate (SFR) that is presented in \autoref{table:NED},  has been obtained by \citet{2009ApJ...706..599L}.
\par This work supplements our previous multi-wavelength study of SNRs in nearby galaxies, which showed evidence for differences in SNRs between spiral and irregular galaxies.  More specifically, it is seen that more luminous X-ray emitting SNRs tend to be preferentially associated with irregular galaxies (see Fig.12 of \citealt{2010ApJ...725..842L}) which is attributed to different ISM density between the two types of galaxies.



\begin{table*}

\caption{Galaxies Properties.}
\hspace{-0.5cm}
\begin{tabular}{l ccccccc}
\hline
\small Galaxy & \small Distance$^{*}$ &\small Size$^{**}$ &\small RA &\small Dec &\small Metallicity &\small SFR(H$\rm \alpha$)& Chandra data\\
    &      &   & \small(J2000) &\small (J2000) & \\
    & \small (Mpc)    &\small (arcmin) & \small hh:mm:ss &\small  dd:mm:ss & & $\rm{M_{\odot}\,yr^{-1}}$& total exposure time$^{***}$ (ks)\\
\hline 
\small NGC\,45   &\small \hfil 6.79  &\small \hfil  8.5, 5.9   &\small  00:14:03.99  &\small \hfil -23:10:55.5  &\small \hfil  8.51 &\small \hfil  0.39&\small \hfil 65.06\\

\small NGC\,55   &\small \hfil 1.99  &\small \hfil  32.4, 5.6  &\small 00:14:53.60  &\small \hfil -39:11:47.9  &\small \hfil  8.54 &\small \hfil  0.47 &\small \hfil 68.93\\

\small NGC\,1313 &\small \hfil 4.15  &\small \hfil  9.1, 6.9   &\small 03:18:16.05  &\small \hfil -66:29:53.7   &\hfil  8.62 &\hfil  0.67&\small \hfil 68.98\\

\small NGC\,7793 &\small \hfil 3.70  &\small \hfil  9.3, 6.3   &\small 23:57:49.83  &\small \hfil -32:35:27.7   &\small \hfil  8.54 &\small \hfil  0.51 &\small \hfil 2190.26\\
\hline
\end{tabular}
\label{table:NED}
\begin{tablenotes}
\item Col (1): The name of the galaxies; col (2) the distance of the galaxies; col (3) the size of the galaxies; col (4) and col (5): the Right Ascension and Declination of the galaxies; col (6): the $\rm{[12+log(O/H)]}$ metallicity of the galaxies; col (7): the extinction corrected H$\rm \alpha$ SFR. (\citealt{2009ApJ...706..599L}).\\
  $^{*}$The distance and the size characteristics are taken from the NED (NASA/IPAC Extragalactic Database; \small{http://ned.ipac.caltech.edu/}).\\
$^{**}$Major and minor axis.\\
$^{***}$ The exposure time does not include observations in sub-array mode.
\end{tablenotes}
\end{table*}

\par All four galaxies are known to host SNRs identified in optical, radio, or X-ray wavelengths. However, the available studies are not systematic, not allowing the investigation of their populations in detail. From our sample, only NGC 7793 has been studied in optical wavelengths. \citet{Blair1997} detected and spectroscopically confirmed 27 SNRs, with H$\rm \alpha$ fluxes ranging from $\rm 6.7\times 10^{-16}$ to $\rm 3.6\times 10^{-14}\, erg\, cm^{-2}\, s^{-1}$. Two of them have been also confirmed by \citet{2020A&A...635A.134D}. \cite{2011AJ....142...20P,2002ApJ...565..966P} presented 7 radio SNRs, 2 of which coincide with optical SNRs, while the rest of them are new identification. More recently, \citet{2014ApSS.353..603G} presented a catalogue of 14 radio SNRs  that includes 5 of the 7 aforementioned radio SNRs. Two of them coincide with optical SNRs from the work of \citet{Blair1997}.


\citeauthor{1980A&AS...40...67D} (\citeyear{1980A&AS...40...67D}) looked for optical SNRs in NGC 45 but without any success. However, a candidate SNR in radio with no X-ray counterpart and a candidate one in X-rays are reported in the work of \citeauthor{2015AJ....150...91P} (\citeyear{2015AJ....150...91P}). 
NGC 55 has given 6 radio candidate SNRs that are reported to have also X-ray emission  (\citealt{2013Ap&SS.347..159O}).  Five SNRs have been detected in the X-ray band (\citealt{2006MNRAS.370...25S}) and 2 of them have also emission in radio (\citealt{Hummel1986}).
In addition, 13 more X-ray SNRs have been reported in the study of \citet{2015AJ....150...94B}.
In NGC 1313, a probable young SNR around the supernova SN 1978K has been identified in the X-rays  (\citealt{smith2007}, \citealt{1995ApJ...446..177C}; \citealt{1996ApJ...456..187S}; \citealt{1994PASJ...46L.115P}) and in radio (\citealt{1994A&A...285..687A}). \citet{1995ApJ...446..177C} presents also 4 probable SNRs in X-rays.

\begin{table*}
\caption{Observation details}

\begin{threeparttable}

\begin{tabular}{l c c c l c c}
\hline
Filters & RA & Dec &  Exposure Time  & Exposures & PSF  & Date   \\ 
        & (J2000) & (J2000) &  (sec)  &  & (arcsec$^*$) &   \\ 
\hline 
\hline
\multicolumn{7}{c}{NGC\,45}\\
\hline 
H$\rm{\alpha}$ + [\ion{N}{II}]  &  00:14:03.48  &\hfil -23:10:56.20 &  3600 & 5 $\times$ 720$\rm s$ & 0.95 & 17 Nov 2011\\

[\ion{S}{II}]  &  00:14:04.11  &\hfil -23:10:37.80  &  7200  &  5 $\times$ 1440$\rm s$ &  1.30 & 17 Nov 2011\\

R  &  00:14:05.10  &\hfil -23:10:07.70 & 600 & 5 $\times$ 120$\rm s$ & 1.30 & 17 Nov 2011\\

\hline
\multicolumn{7}{c}{NGC\,55}\\
\hline 
H$\rm{\alpha}$ + [\ion{N}{II}] &  00:14:53.95 &\hfil -39:11:46.39  & 3600 &  5 $\times$ 720$\rm s$   & 1.03 &  16 Nov 2011 \\

[\ion{S}{II}]  &  00:14:53.53  &\hfil -39:11:47.60 & 7200 &  5 $\times$ 1440$\rm s$ & 1.22 & 16 Nov 2011\\

R  &  00:14:53.67  &\hfil -39:11:48.50  & 600 &  5 $\times$ 120$\rm s$ & 1.32 & 16 Nov 2011\\

\hline
\multicolumn{7}{c}{NGC\,1313}\\
\hline 
H$\rm{\alpha}$ + [\ion{N}{II}]  &  03:18:16.75  &\hfil -66:29:52.00 & 3600  & 5 $\times$ 720$\rm s$ & 0.95 & 15 Nov 2011\\

[\ion{S}{II}]  &  03:18:17.60  &\hfil -66:29:54.49 & 5760 & 4 $\times$ 1440$\rm s$ & 1.30 & 16 Nov 2011\\

R  &  03:18:16.01  &\hfil -66:29:53.99 & 600  & 5 $\times$ 120$\rm s$ & 0.92 & 15 Nov 2011\\

\hline
\multicolumn{7}{c}{NGC\,7793}\\
\hline
H$\rm{\alpha}$ + [\ion{N}{II}]  &  23:57:49.82  &\hfil -32:35:28.10 & 3600 & 5 $\times$ 720$\rm s$ & 0.95 & 15 Nov 2011\\

[\ion{S}{II}]  &  23:57:49.83  &\hfil -32:35:28.20 & 7200  & 5 $\times$ 1440$\rm s$ & 0.97 & 15 Nov 2011\\

R  &  23:57:49.78  &\hfil -32:35:28.20 & 600  & 5 $\times$ 120$\rm s$ & 1.05 & 15 Nov 2011\\

\hline
\end{tabular}
\begin{tablenotes}
\item Col (1): The filters that were used; col (2) and col (3): the RA and Dec of each exposure; col (4) the exposure time: col (5): the number of the exposures; col (6): the PSF of the obtained images; col (7): the date of each observation.
 \\
$^*$ It is the FWHM of the PSF.
\end{tablenotes}
\end{threeparttable}
\label{table:logs}
\end{table*}

\section{Observations and Data reduction} \label{Observations and Data reduction}
The data used in this work were obtained with the 4-m Blanco telescope at CTIO (Chile) on November 15-17, 2011. We used the {$36\arcmin \times 36\arcmin$} Mosaic II CCD imager which consists of 8 CCDs ($2048 \times 2048$ SITe each) with a pixel scale of $0.27\arcsec/\rm{pixel}$. We used the narrow band H$\rm{\alpha}$ + [\ion{N}{II}]  and [\ion{S}{II}]  filters and a broadband continuum R filter. For the H$\rm{\alpha}$ + [\ion{N}{II}] filter, the central wavelength is 6563{\AA} with a FWHM of 80 {\AA}, for the [\ion{S}{II}] filter the central wavelength is 6725 {\AA} with a FWHM of 80 {\AA}, and for the R continuum, the central wavelength is 6440 {\AA} with a FWHM of 1510 {\AA}. 

\par The total exposure time for each galaxy was 3600 sec for the H$\rm{\alpha}$ + [\ion{N}{II}], 7200 for the [\ion{S}{II}], and 600 sec for the R-continuum filters. The integration time for each observation was split into 5 shorter exposures. For the larger galaxies, the pointing for each of the 5 observations was shifted slightly  to cover the chip gaps. This dither procedure ensures uniform coverage of the galaxy and efficient removal of cosmic rays. Detailed information on the observations for each galaxy is given on \autoref{table:logs}. 

\par For the reduction of the mosaic images, we used the \texttt{mscred} package of IRAF{\footnote{http://ast.noao.edu/data/software}}  (Image Reduction and Analysis Facility; \citealt{1993ASPC...52..173T,1986SPIE..627..733T}). Bias and flat-field corrections were performed on all images; astrometric calibrations were applied on the reduced CCD images using the 2MASS catalog and a fourth order polynomial to account for distortions at the edges of the images. 
The individual exposures for each object and each filter were registered and median combined with SWarp\footnote{https://www.astromatic.net/software/swarp}

\section{Detection and Photometry}  \label{Detection and Photometry}
The first step in our analysis was the detection of discrete sources in each galaxy. To do this, we used the program SExtractor\footnote{https://www.astromatic.net/software/sextractor} (\citealt{sextractor}). SExractor was chosen because in comparison to similar tools, it is more effective in the detection of fainter objects located in dense star fields, as well as spatially varying backgrounds, with significant structure (\citealt{2013PASP..125...68A}). We ran SExtractor setting the following parameters for the  H$\rm \alpha$ + [\ion{S}{II}] image: a) detection threshold at 2.5$\rm \sigma$  above the background, b) the minimum number of source pixels equal to 5, and c) the background mesh size at 5 pixels in order to account for small-scale background variations.
From the detected sources in each image, we kept only those within the optical outline of the galaxy in H$\rm\alpha$. This region is defined as the region for which the intensity in H$\rm \alpha$ is 3$\sigma$ above the background.

\par The next step was the photometric analysis of the sources which was performed using the package \texttt{phot} of IRAF. The intensity of the detected sources was measured by means of a curve of growth analysis (i.e. we measure the source intensity in apertures of increasing radius and adopting the one maximising the signal to noise ratio).  The optimal aperture  radius for the majority of sources is 4.0 pixels  corresponding to 35.2 pc, 10.3 pc, 21.5 pc, and 19.2 pc, for  NGC 45, NGC 55, NGC 1313, and NGC 7793 respectively.

\par In order to subtract the star-light continuum from the detected sources, for each galaxy we selected stars of moderate intensity  and for each of them we calculated the intensity ratios $\rm{(H\alpha + [\ion{N}{II}])/R}$ and $\rm{[\ion{S}{II}]/R}$  where H$\rm{\alpha}$, [\ion{S}{II}], and R are the measured intensities in the  H$\alpha$, [\ion{S}{II}] and R-continuum images respectively. The mode of these ratios was used as a scaling factor for the R-continuum intensity before subtracting it from the measured  H$\rm{\alpha + [\ion{N}{II}]}$ and [\ion{S}{II}] intensities. 

An additional correction needed to obtain the net H$\rm \alpha$ flux is to subtract the contribution of the [\ion{N}{II}]$\rm \lambda \lambda$6548,6584\AA\,  lines .
 We adopted a $\rm{([\ion{N}{II}] \lambda \lambda 6548, 6584)/H\alpha}$  flux ratio of 0.27, based on emission-line measurements of spectroscopically identified SNRs of galaxies with similar metallicities of our sample (\citealt{Long2018},  \citealt{2013MNRAS.429..189L}, \citealt{Blair1997}).



 To examine possible contamination by \ion{H}{II} regions resulting from over-subtraction of the [\ion{N}{II}] contamination, we considered the spectroscopic [\ion{N}{II}]/H$\rm\alpha$ ratio of \ion{H}{II} regions in  NGC 1232 (\citealt{2020A&A...642A.203L}), NGC 300 and NGC 7793 (\citealt{Blair1997}), NGC 7793 (\citealt{2010MNRAS.405.2737B}), M31 (\citealt{2012MNRAS.427.1463Z}) and M33 (\citealt{2017ApJ...842...97L}). From the average values of the ratios reported in each of those works, the lowest is 0.19. 
  We find that by adopting the ratio [\ion{N}{II}]/H$\rm\alpha$ = 0.27 
  there are $\sim$ 29\% more  \ion{H}{II} regions (for [\ion{S}{II}]/H$\rm\alpha$ ratio 3$\rm \sigma$ above the 0.4 threshold) than when we consider the ratio [\ion{N}{II}]/H$\rm\alpha$ = 0.19  for the \ion{H}{II} regions. This is a rather conservative fraction since we consider the lowest [\ion{N}{II}]/H$\rm\alpha$ ratio of the \ion{H}{II} region sample (the mean value of the [\ion{N}{II}]/H$\rm \alpha$ ratio of the rest of the studies ranges from 0.2-0.38).
  In general, it is very difficult to have zero contamination by  \ion{H}{II} regions in an SNR photometric sample, especially because the contamination by [\ion{N}{II}] in the H$\rm\alpha$ flux cannot be predicted exactly for each source. However, a spectroscopic SNR sample can also suffer from contamination by \ion{H}{II} regions  because there are  \ion{H}{II} regions that may have [\ion{S}{II}]/H$\rm\alpha$ >  0.4. This can be alleviated by using 
  diagnostics that combine more line ratios (\citealt{2020MNRAS.491..889K}).
 
 %


\par In order to flux calibrate the detected sources, we used observations of standard stars from the catalogue of \citet{Massey}. We obtained photometry for these stars at different airmasses and we performed a linear fit of their instrumental magnitude against the airmass $\rm \chi$: $\rm {m = m_{inst} - k\chi + b}$ where $\rm m$ is the calibrated magnitude, and $\rm k$ and $\rm b$ the slope and the intercept that are to be calculated. 
Hence we have:
\\
\\
 \hspace{0.5cm} $\rm {m_v = -2.5log(f_{\nu}) - 48.9 = -2.5log(count\, rate) - kx + b}$ \\ 
\\
where $\nu$ is the central frequency of the filter that we use. Finally, in order to convert flux density to flux, we also multiply with the FWHM (Full Width at Half Maximum) of the filter. 

\par At this point, we have the continuum subtracted H$\rm{\alpha}$ and [\ion{S}{II}] flux of every detected source. In order to find which of them are SNRs, we followed a totally automated method. 
From the full sample of sources, we first consider those with H$\rm{\alpha}$ flux above the 3$\rm{\sigma}$ level with respect to the local background. Then,  the SNRs are identified by applying the standard [\ion{S}{II}]/H$\rm{\alpha}$ > 0.4 criterion (\citealt{{Mathewson&Clarke}}),  considering  sources with [\ion{S}{II}]/H$\rm{\alpha}$ > 0.4  at the 3$\rm{\sigma}$ level (candidate SNRs) and at the 2$\rm{\sigma}$ (possible candidate SNRs). To ensure the reliability of the sources identified in this automated way, we visually inspected each one of them, on the  H$\rm{\alpha}$, [\ion{S}{II}], and R images. This showed that some of them found to be parts of larger structures, for example rings, large bubbles and filaments. These extended sources are not considered for further analysis. All other candidate SNRs are reliable identifications, based on their appearance on the [\ion{S}{II}] and H$\rm  \alpha$ images, ensuring the reliability of the automated method.

\subsection{Calculation of Incompleteness} \label{incomp}
During the detection process, a fraction of the sources is not detected due to their faintness in combination to their environment and stochastic effects. For this reason, our sample (as any observational sample) is characterized by incompleteness which depends on the H$\rm{\alpha}$ local background, the detection method, and sample selection effects. In order to evaluate this incompleteness, we followed the standard approach of placing artificial objects on our images (with the  IRAF task \texttt{addstar}) using their measured PSF (determined with the IRAF package \texttt{PSF}) and in a wide range of magnitudes, covering the full brightness range of the observed objects in each filter.

\par Then we perform the detection exactly as we did for the actual data.  The process of adding and recovering artificial sources is repeated multiple  times  to improve our statistics. In every iteration we took special care to include at most as many artificial sources as objects observed in the actual data, in order to avoid increasing the source confusion. Then, we performed aperture photometry on the artificial objects, the same way as in the actual data, keeping only the sources that are 3$\rm{\sigma}$ above the background in H$\rm{\alpha}$, then the sources with [\ion{S}{II}]/H$\rm{\alpha}$ > 0.4, and finally the sources for which the ratio is 3$\rm{\sigma}$ above 0.4. This results in three samples: an H$\rm{\alpha}$ sample, a [\ion{S}{II}]/H$\rm{\alpha}$ selected sample, and a "secure" [\ion{S}{II}]/H$\rm{\alpha}$ sample ([\ion{S}{II}]/H$\rm{\alpha}$ > 0.4 at 3$\rm{\sigma}$ significance).

\par  In order to create a 2D incompleteness map, we divide the input artificial star magnitude range of  the (H$\rm \alpha$, \ion{S}{II}) plane, in 2 dimensional bins, and we calculate the fraction of input and detected objects (after applying all relevant selection criteria) for each ($\rm L_{H\alpha}, L_{[S\,II]}$) bin. This comparison is performed on the continuum-subtracted fluxes: the continuum subtraction is achieved by subtracting the R-band continuum flux using the same scaling factor as for the real data, since the artificial objects are placed on the same background as the actual data. So now, we have the fraction of detected/input objects (incompleteness) as a function of their H$\rm{\alpha}$ and [\ion{S}{II}] intensities. 

\par Because of the varying diffuse emission surface brightness, the incompleteness varies within each galaxy. To account for this variation, we calculate the incompleteness for four different background regimes for each galaxy. These are selected to trace the full range of backgrounds in which sources are detected in each galaxy. Indicatively, in \autoref{fig:7793_im} we show the different regions with contours for NGC 7793.


\par In order to estimate the false positives (the sources that have been measured to have [\ion{S}{II}]/H$\rm{\alpha}$ > 0.4 at the 3$\rm \sigma$ level while in reality they do not), from the initial artificial objects we created two sub-samples, one for sources with input [\ion{S}{II}]/H$\rm{\alpha}$ ratio between 0.2 and 0.3, and one for input ratio between 0.3 and 0.4. Then, we examined what fraction of sources detected with ratio > 0.4 belongs to these two sub-samples.

\section{Results} \label{results}
\subsection{Candidate SNRs}

\par In total, we detected 42 candidate SNRs (with [\ion{S}{II}]/H$\rm{\alpha}$ ratio 3$\rm{\sigma}$ above the 0.4 threshold, 30 of which are new identifications)  and 54 new possible candidate SNRs (with [\ion{S}{II}]/H$\rm{\alpha}$ ratio 2$\rm{\sigma}$ above the 0.4 threshold). In addition, some of the knots that present high excitation ([\ion{S}{II}]/H$\rm{\alpha}$ > 0.4), in both 3$\rm{\sigma}$ and 2$\rm{\sigma}$ samples, are part of larger structures (i.e shells, bubbles etc, with sizes from $\sim$ 40 pc $-$ $\sim$ 320 pc). In total, we identify 21 such structures, with those with small radii (i.e. $ < $ 100 pc) probably being SNRs. These objects are not considered in our candidate SNR samples, but they are reported for completeness and follow-up observations.

\begin{figure*}
\centering
 \includegraphics[width=0.7\textwidth]{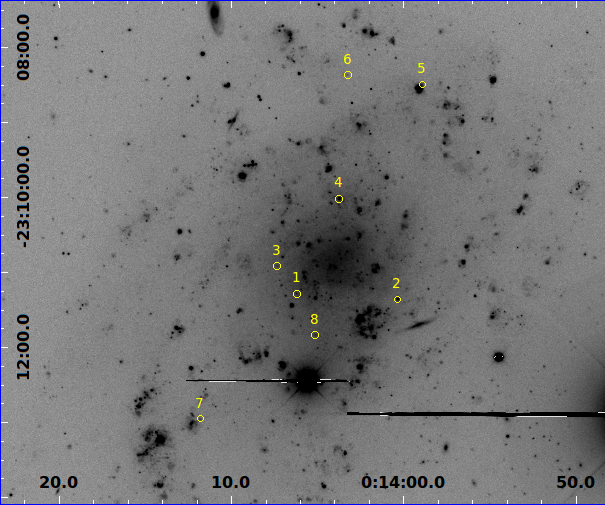}
\caption{\label{fig:45_im}{ \textbf{NGC\,45}: The candidate SNRs (yellow circles) overlaid on the H$\alpha$ + [\ion{N}{II}] image. In this galaxy we identified 8 candidate SNRs with [\ion{S}{II}]/H$\rm{\alpha}$ ratio 3$\rm{\sigma}$ above 0.4. The flux of the faintest source in H$\rm{\alpha}$ is 3.8$\times 10^{-16}\rm{erg\,cm^{-2}\,s^{-1}}$.}}
\end{figure*}

\begin{figure*}
\centering
 \includegraphics[width=0.9\textwidth]{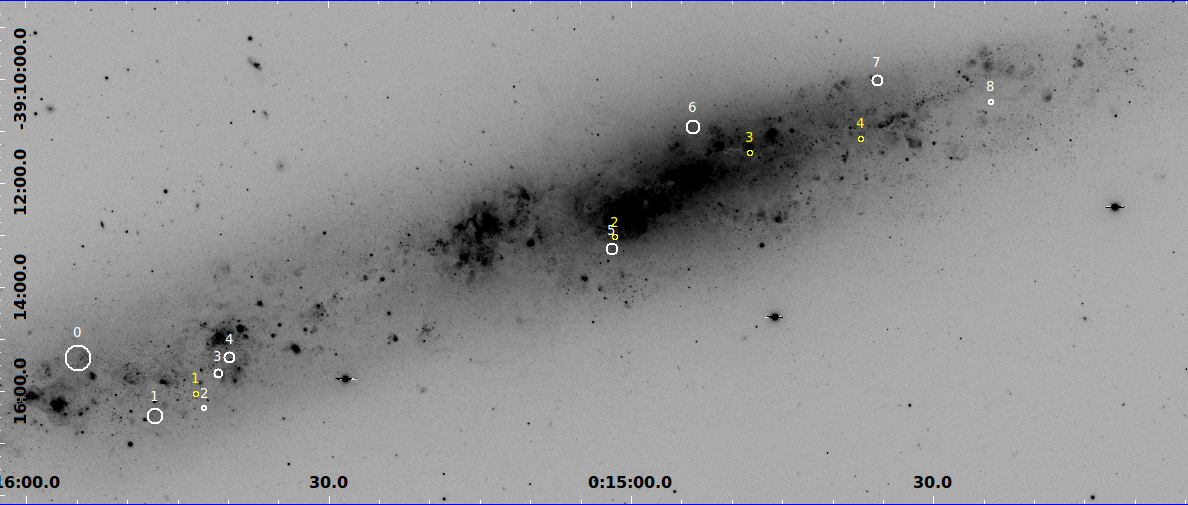}
\caption{\label{fig:55_im} {\textbf{NGC\,55}: The candidate SNRs (yellow circles) and the larger structures (white circles) overlaid on the H$\alpha$ + [\ion{N}{II}] image. In this galaxy we identified 4 candidate SNRs with [\ion{S}{II}]/H$\rm{\alpha}$ ratio 3$\rm{\sigma}$ above 0.4. The flux of the faintest source in H$\rm{\alpha}$ is 7.59$\times 10^{-16}\rm{erg\,cm^{-2}\,s^{-1}}$.}}
\end{figure*}

\begin{figure*}
\centering
 \includegraphics[width=0.7\textwidth]{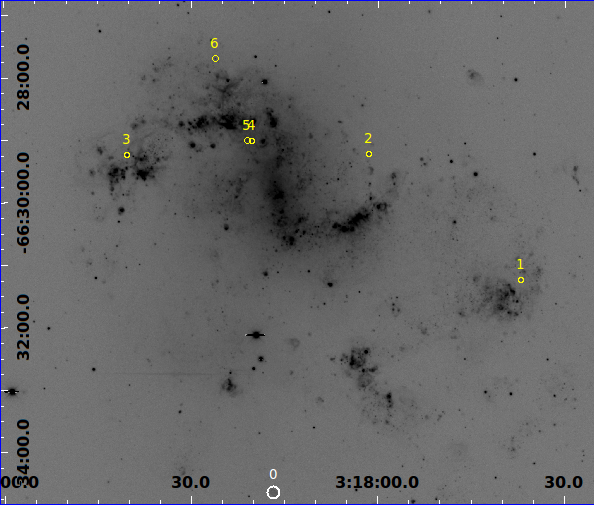}
\caption{\label{fig::1313_im}{\textbf{NGC\,1313}: The candidate SNRs (yellow circles) and the larger structures (white circles) overlaid on the H$\alpha$ + [\ion{N}{II}] image. In this galaxy we identified 6 candidate SNRs with [\ion{S}{II}]/H$\rm{\alpha}$ ratio 3$\rm{\sigma}$ above 0.4. The flux of the faintest source in H$\rm{\alpha}$ is 6.8$\times 10^{-16}\rm{erg\,cm^{-2}\,s^{-1}}$}}
\end{figure*}

\begin{figure*}
\centering
 \includegraphics[width=0.8\textwidth]{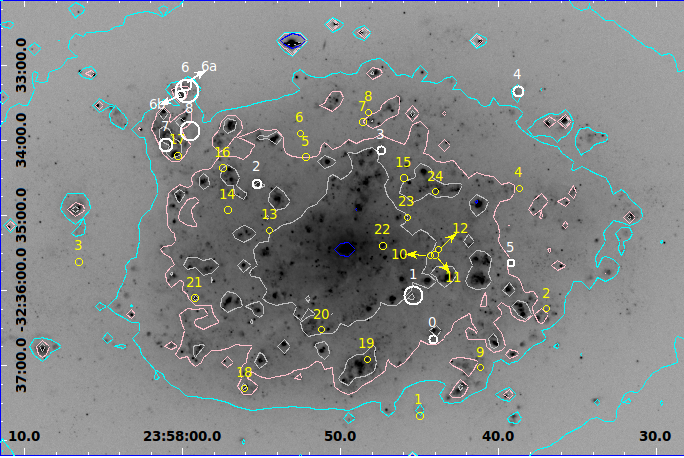}
\caption{\label{fig:7793_im} {\textbf{NGC\,7793}: The candidate SNRs (yellow circles) and the larger structures (white circles) overlaid on the H$\alpha$ + [\ion{N}{II}] image. In this galaxy we identified 24 candidate SNRs with [\ion{S}{II}]/H$\rm{\alpha}$ ratio 3$\rm{\sigma}$ above 0.4. The flux of the faintest source in H$\rm{\alpha}$ is 3.4$\times 10^{-16}\rm{erg\,cm^{-2}\,s^{-1}}$. The contours show the different background regions for which we calculated the incompleteness. Cyan color is for regions with background 500-600 counts, pink color is for background 600-800 counts, grey color is for background 800-1100 counts, and blue color is for background 1100-3000 counts.}}
\end{figure*}

 In \autoref{table:all_SNRs_total}, we present  the candidate SNRs (> 3$\rm \sigma$). The first column gives the ID of the source, the second and the third  columns give the RA and Dec coordinates (in J2000), the fourth and fifth columns give the H$\rm{\alpha}$ and [\ion{S}{II}] fluxes with their uncertainties, and the sixth column gives the [\ion{S}{II}]/H$\rm{\alpha}$ ratio with its uncertainty. Figures \ref{fig:45_im} - \ref{fig:7793_im} show the location of the candidate SNRs (yellow circles) overlaid on the H$\rm \alpha$ images for each galaxy.

\par We see that SNRs are generally located in star-forming regions as demonstrated by their strong H$\rm \alpha$ emission. In the case of NGC 45, NGC 1313 and NGC 7793  we see a deficit of sources in their central region. This is the result of high background in these regions, resulting in significant incompleteness.
 The possible candidate SNRs (> 2$\rm \sigma$) are presented in the \Cref{SNR_2s}.

\par
\begin{figure*}

\begin{minipage}{0.33\textwidth}
\includegraphics[width=1\textwidth]{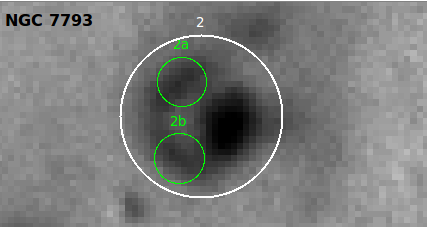}
\end{minipage}\hfill
\begin{minipage}{0.33\textwidth}
\includegraphics[width=1\textwidth]{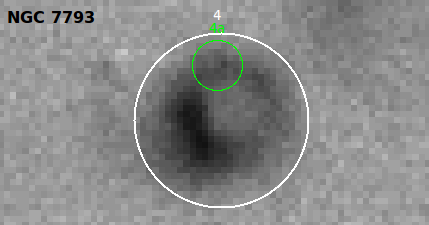}
\end{minipage}\hfill
\begin{minipage}{0.33\textwidth}
\includegraphics[width=1\textwidth]{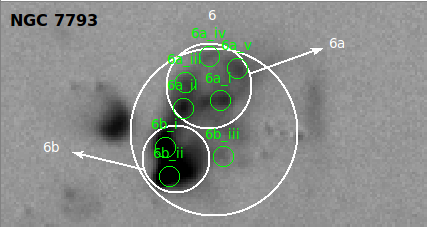}
\end{minipage}

\begin{minipage}{0.33\textwidth}
\includegraphics[width=1\textwidth]{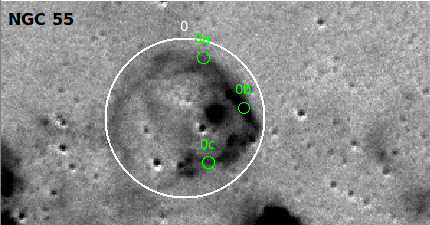}
\end{minipage}\hfill
\begin{minipage}{0.33\textwidth}
\includegraphics[width=1\textwidth]{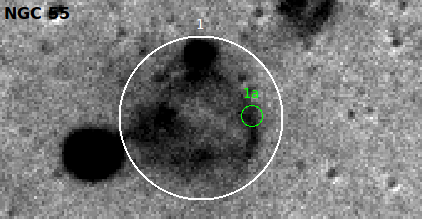}
\end{minipage}\hfill
\begin{minipage}{0.33\textwidth}
\includegraphics[width=1\textwidth]{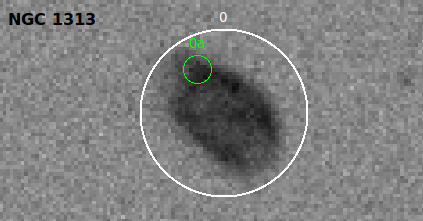}
\end{minipage}
        \caption[]
        {Indicative postage-stamp continuum  subtracted H$\alpha$ + [\ion{N}{II}] images of some of the larger structures presented in  \autoref{table:SNRs_total_rings}. The white circles indicate  the outline of the structures, while the green circles show the location of the detected knots.}     
        \label{fig:structures}
\end{figure*}

  The white circles in figures \ref{fig:45_im} - \ref{fig:7793_im} are the larger structures, with their size indicating their physical size. These structures along with the respective detected knots are presented in \autoref{table:SNRs_total_rings}, where we also give their size. We show some of them in \autoref{fig:structures}.

\onecolumn

\begin{ThreePartTable}
\begin{longtable}{cccccc}
\caption{Candidate SNRs}\\
\hline 
ID & RA & Dec & $\rm{F_{H\alpha}} \pm \rm{\delta F_{H\alpha}}$ & $\rm{F_{[\ion{S}{II}]}} \pm \rm{\delta F_{[\ion{S}{II}]}}$ & $\frac{\rm{F_{[\ion{S}{II}]}}}{\rm{F_{H\alpha}}} \pm $ ($\rm{\delta}\frac{\rm{F_{[\ion{S}{II}]}}}{\rm{F_{H\alpha}}} $)\\ 
 & (J2000) & (J2000) &  &  &  \\
 & hh:mm:ss & dd:mm:ss & ($ \rm 10^{-16}\, erg\,s^{-1}\,cm^{-2}$) & ($\rm  10^{-16}\, erg\,s^{-1}\,cm^{-2}$) &  \\ 
\hline
\multicolumn{6}{c}{NGC 45\,($\rm 3\sigma$)}\\		
\hline
1 & 00:14:06.2 & -23:11:17.0 &   5.91  $\pm$   0.34 &   4.23  $\pm$   0.41 &  0.72 $\pm$ 0.08 \\
2 & 00:14:00.4 & -23:11:21.6 &   19.20 $\pm$   0.55 &   12.70 $\pm$   0.54 &  0.66 $\pm$ 0.03 \\
3 & 00:14:07.3 & -23:10:54.8 &   3.84  $\pm$   0.24 &   3.04  $\pm$   0.26 &  0.79 $\pm$ 0.08 \\
4 & 00:14:03.8 & -23:10:01.0 &   53.70 $\pm$   0.60 &   27.00 $\pm$   0.52 &  0.50 $\pm$ 0.01 \\
5 & 00:14:03.3 & -23:08:22.1 &   11.90 $\pm$   0.59 &   8.14  $\pm$   0.52 &  0.68 $\pm$ 0.06 \\
6 & 00:14:11.8 & -23:12:57.4 &   4.29  $\pm$   0.28 &   2.85  $\pm$   0.26 &  0.66 $\pm$ 0.07 \\
7 & 00:14:05.1 & -23:11:50.0 &   5.50  $\pm$   0.32 &   2.99  $\pm$   0.29 &  0.54 $\pm$ 0.06 \\
8 & 00:13:58.9 & -23:08:29.7 &   4.66  $\pm$   0.28 &   2.77  $\pm$   0.28 &  0.59 $\pm$ 0.07 \\
\hline
\multicolumn{6}{c}{NGC 55\,($\rm 3\sigma$)}\\
\hline
1 & 00:15:43.2 & -39:16:03.0 &   7.59 $\pm$   0.48 &   6.65 $\pm$   0.54 &  0.88 $\pm$ 0.09 \\
2 & 00:15:01.7 & -39:13:03.1 &   10.9 $\pm$   0.62 &   6.94 $\pm$   0.65 &  0.63 $\pm$ 0.07 \\
3 & 00:14:48.2 & -39:11:26.1 &   60.4 $\pm$    1.9 &   42.6 $\pm$    2.0 &  0.71 $\pm$ 0.04 \\
4 & 00:14:37.3 & -39:11:09.8 &   19.1 $\pm$   0.83 &   12.2 $\pm$    0.8 &  0.64 $\pm$ 0.05 \\
\hline
\multicolumn{6}{c}{NGC 1313\,($\rm 3\sigma$)}\\
\hline
1 & 03:17:37.1 & -66:31:13.9 &   9.51 $\pm$   0.37 &   5.30 $\pm$   0.30 &  0.56 $\pm$ 0.04 \\
2 & 03:18:01.5 & -66:29:13.6 &   12.2 $\pm$   0.91 &   8.64 $\pm$   0.75 &  0.71 $\pm$ 0.08 \\
3 & 03:18:40.3 & -66:29:14.3 &   59.9 $\pm$   0.57 &   32.3 $\pm$   0.56 &  0.54 $\pm$ 0.01 \\
4 & 03:18:20.3 & -66:29:00.9 &   6.82 $\pm$   0.64 &   5.06 $\pm$   0.54 &  0.74 $\pm$ 0.10 \\
5 & 03:18:21.0 & -66:29:00.3 &   15.1 $\pm$   0.89 &   10.1 $\pm$   0.64 &  0.67 $\pm$ 0.06 \\
6 & 03:18:26.1 & -66:27:41.8 &   15.8 $\pm$   0.42 &   11.3 $\pm$   0.44 &  0.71 $\pm$ 0.03 \\
\hline
\multicolumn{6}{c}{NGC 7793\,($\rm 3\sigma$)}\\
\hline
1 & 23:57:45.0 & -32:37:40.2 &      16.0 $\pm$    0.63 &      16.0 $\pm$    0.51 &  0.98 $\pm$ 0.05 \\
2 & 23:57:37.0 & -32:36:14.9 &     8.8 $\pm$    0.54 &     7.1 $\pm$    0.55 &  0.81 $\pm$ 0.08 \\
3 & 23:58:06.5 & -32:35:37.0 &      11.0 $\pm$    0.33 &     9.9 $\pm$    0.41 &  0.88 $\pm$ 0.05 \\
4 & 23:57:38.7 & -32:34:38.5 &     6.4 $\pm$    0.48 &     8.1 $\pm$     0.6 &  1.30 $\pm$ 0.10 \\
5 & 23:57:52.2 & -32:34:13.4 &      10.0 $\pm$    0.32 &     6.6 $\pm$    0.43 &  0.66 $\pm$ 0.05 \\
6 & 23:57:52.6 & -32:33:54.4 &      5.0 $\pm$     0.4 &     5.7 $\pm$    0.39 &  1.10 $\pm$  0.10 \\
7 & 23:57:48.6 & -32:33:45.4 &     3.4 $\pm$    0.38 &     3.1 $\pm$    0.39 &  0.90 $\pm$  0.10 \\
8 & 23:57:48.2 & -32:33:37.7 &     5.4 $\pm$    0.49 &     5.4 $\pm$    0.47 &  0.99 $\pm$  0.10 \\
9 & 23:57:41.1 & -32:37:02.1 &      24.0 $\pm$     1.2 &      21.0 $\pm$    0.93 &  0.88 $\pm$ 0.06 \\
10 & 23:57:44.3 & -32:35:32.1 &     6.2 $\pm$    0.44 &     7.3 $\pm$    0.47 &  1.20 $\pm$  0.10 \\
11 & 23:57:44.0 & -32:35:31.7 &     6.6 $\pm$    0.48 &     6.1 $\pm$     0.5 &  0.94 $\pm$  0.10 \\
12 & 23:57:43.8 & -32:35:27.8 &      24.0 $\pm$    0.89 &      23.0 $\pm$     1.3 &  0.95 $\pm$ 0.06 \\
13 & 23:57:54.5 & -32:35:12.2 &      16.0 $\pm$     1.0 &      18.0 $\pm$     1.4 &   1.20 $\pm$  0.10 \\
14 & 23:57:57.2 & -32:34:55.9 &     4.9 $\pm$    0.45 &     5.9 $\pm$    0.56 &   1.20 $\pm$  0.20 \\
15 & 23:57:46.0 & -32:34:30.1 &     6.9 $\pm$    0.92 &     6.7 $\pm$    0.72 &  0.97 $\pm$  0.20 \\
16 & 23:57:57.5 & -32:34:21.9 &      27.0 $\pm$     1.6 &      22.0 $\pm$     1.2 &  0.82 $\pm$ 0.07 \\
17 & 23:58:00.3 & -32:34:12.2 &     100.0 $\pm$     2.5 &      53.0 $\pm$     2.5 &  0.51 $\pm$ 0.03 \\
18 & 23:57:56.1 & -32:37:18.5 &      56.0 $\pm$     1.9 &      36.0 $\pm$     2.2 &  0.64 $\pm$ 0.05 \\
19 & 23:57:48.3 & -32:36:55.2 &      73.0 $\pm$     2.3 &      52.0 $\pm$     1.7 &   0.70 $\pm$ 0.03 \\
20 & 23:57:51.2 & -32:36:31.7 &      38.0 $\pm$     1.9 &      48.0 $\pm$     2.2 &   1.30 $\pm$ 0.09 \\
21 & 23:57:59.2 & -32:36:06.0 &     140.0 $\pm$     2.4 &      83.0 $\pm$     2.5 &  0.61 $\pm$ 0.02 \\
22 & 23:57:47.3 & -32:35:23.9 &     110.0 $\pm$     3.0 &      88.0 $\pm$     3.0 &  0.82 $\pm$ 0.04 \\
23 & 23:57:45.8 & -32:35:01.7 &      33.0 $\pm$    0.77 &      33.0 $\pm$    0.84 &  1.00 $\pm$ 0.03 \\
24 & 23:57:44.0 & -32:34:41.3 &      50.0 $\pm$     1.8 &      27.0 $\pm$     1.8 &  0.54 $\pm$ 0.04 \\
\hline
\label{table:all_SNRs_total}
\end{longtable}

{(The sample of candidate SNRs with significance lower than $\rm 3\sigma$ is presented in \autoref{table:SNRs_total_2s}.)}
\end{ThreePartTable}

\clearpage
\begin{ThreePartTable}
\footnotesize
\begin{longtable}{lccllll}

\caption{Candidate SNRs - Large structures}\\
\hline 
ID & RA (J2000) & Dec (J2000) & $\rm{F_{H\alpha}} \pm \rm{\delta F_{H\alpha}}$ & $\rm{F_{[\ion{S}{II}]}} \pm \rm{\delta F_{[\ion{S}{II}]}}$ & $\frac{\rm{F_{[\ion{S}{II}]}}}{\rm{F_{H\alpha}}} \pm $ ($\rm{\delta}\frac{\rm{F_{[\ion{S}{II}]}}}{\rm{F_{H\alpha}}} $) & $\rm Size$\\ 
 & hh:mm:ss & dd:mm:ss & ($ \rm 10^{-16}\, erg\,s^{-1}\,cm^{-2}$) & ($\rm  10^{-16}\, erg\,s^{-1}\,cm^{-2}$) & &(pc)  \\ 
\hline
\multicolumn{6}{c}{NGC 7793}\\
\hline
\bfseries{0}  & \bfseries{23:57:44.1} & \bfseries{-32:36:39.1} & \bfseries{131.5   $\pm$   4.1} &\bfseries{ 93.5  $\pm$   4.5 } &  \bfseries{0.71 $\pm$   0.02} & \bfseries{101} \\
0a & 23:57:44.0 & -32:36:38.5 &  7.1 $\pm$   0.56 &   5.2 $\pm$   0.69 &  0.74 $\pm$ 0.1 \\
\hline

\hline
\bfseries{1}  & \bfseries{23:57:45.4} & \bfseries{-32:36:03.4} & \bfseries{374.9   $\pm$  13.5} & \bfseries{119.7 $\pm$  17.5} &  \bfseries{0.32 $\pm$   0.01} & \bfseries{218} \\
1a & 23:57:45.4 & -32:35:58.3 &     5.7 $\pm$     0.5 &     4.7 $\pm$    0.75 &  0.82 $\pm$  0.1 \\
\hline
\bfseries{2}  & \bfseries{23:57:55.3} & \bfseries{-32:34:34.4} & \bfseries{214.0   $\pm$   4.7} & \bfseries{125.3 $\pm$   4.7 }&  \bfseries{0.59 $\pm$   0.01} & \bfseries{ 93} \\
2a & 23:57:55.4 & -32:34:32.9 &      12 $\pm$     1.7 &      12 $\pm$     1.7 &  0.97 $\pm$  0.2 \\
2b & 23:57:55.4 & -32:34:36.2 &      31 $\pm$     1.3 &      26 $\pm$       2 &  0.83 $\pm$ 0.07 \\
\hline

\bfseries{3}  & \bfseries{23:57:47.5} & \bfseries{-32:34:08.2} & \bfseries{68.3    $\pm$   3.5} & \bfseries{23.8  $\pm$   3.0} &  \bfseries{0.35 $\pm$   0.02} & \bfseries{ 89 }\\
3a & 23:57:47.4 & -32:34:08.4 &     6.8 $\pm$     1.6 &     7.8 $\pm$     1.3 &   1.1 $\pm$  0.3 \\
\hline
\bfseries{4}  & \bfseries{23:57:38.8} & \bfseries{-32:33:20.8} & \bfseries{99.8    $\pm$   1.8} & \bfseries{42.7  $\pm$   2.0} &  \bfseries{0.43 $\pm$   0.01} & \bfseries{101} \\
4a & 23:57:38.8 & -32:33:18.5 &      12 $\pm$     1.3 &     8.4 $\pm$    0.76 &  0.73 $\pm$  0.1 \\
\hline

\bfseries{5}  & \bfseries{23:57:39.2} & \bfseries{-32:35:38.1} & \bfseries{58.6    $\pm$   1.1} & \bfseries{52.4  $\pm$   1.1} &  \bfseries{0.89 $\pm$   0.02} &  \bfseries{71} \\
5a & 23:57:39.2 & -32:35:36.7 &      16 $\pm$     1.1 &     9.1 $\pm$     1.1 &  0.59 $\pm$ 0.08 \\
5b & 23:57:39.2 & -32:35:39.2 &     9.8 $\pm$    0.69 &      11 $\pm$    0.54 &   1.1 $\pm$ 0.09 \\
\hline
\bfseries{6}  & \bfseries{23:57:59.8} & \bfseries{-32:33:20.5} & \bfseries{1199.0  $\pm$   8.7} & \bfseries{786.9 $\pm$   9.9} &  \bfseries{0.66 $\pm$   0.00} & \bfseries{322} \\
6a & 23:57:59.8 & -32:33:15.4 & 338.4   $\pm$   7.8 & 213.3 $\pm$   6.0 &  0.63 $\pm$   0.01 & 151 \\
6a\_i & 23:57:59.7 & -32:33:17.1 &      16.0 $\pm$    1.0 &      11 $\pm$    0.75 &  0.73 $\pm$ 0.07 \\
6a\_ii & 23:58:00.0 & -32:33:17.9 &      31.0 $\pm$     2.3 &      24 $\pm$     1.8 &  0.8 $\pm$ 0.08 \\
6a\_iii & 23:58:00.0 & -32:33:15.2 &      32.0 $\pm$     1.8 &      22 $\pm$     1.4 &  0.7 $\pm$ 0.06 \\
6a\_iv & 23:57:59.8 & -32:33:12.3 &      10.0 $\pm$    0.8 &     6.2 $\pm$    0.39 &   0.6 $\pm$ 0.06 \\
6a\_v & 23:57:59.6 & -32:33:13.6 &      14.0 $\pm$     1.1 &      11 $\pm$     1.1 &  0.81 $\pm$  0.1 \\
6b & 23:58:00.1 & -32:33:23.0 & 690.3   $\pm$   4.3 & 485.0 $\pm$   4.1 &  0.70 $\pm$   0.00 & 135 \\
6b\_i & 23:58:00.2 & -32:33:22.1 &     100 $\pm$     3.7 &      66 $\pm$     3.2 &  0.65 $\pm$ 0.04 \\
6b\_ii & 23:58:00.1 & -32:33:25.2 &     130 $\pm$     5.6 &     120 $\pm$     4.3 &   0.9 $\pm$ 0.05 \\
6b\_iii & 23:57:59.7 & -32:33:23.2 &     4.8 $\pm$    0.68 &       4 $\pm$    0.62 &  0.83 $\pm$  0.2 \\
\hline

\bfseries{7}  & \bfseries{23:58:01.1} & \bfseries{-32:34:03.8} & \bfseries{242.6   $\pm$  13.8} & \bfseries{36.5  $\pm$   9.6} &  \bfseries{0.15 $\pm$   0.01} & \bfseries{169} \\
7a & 23:58:01.3 & -32:34:05.6 &     9.3 $\pm$     1.3 &     7.2 $\pm$    0.72 &  0.78 $\pm$  0.1 \\

\hline
\bfseries{8}  & \bfseries{23:57:59.5} & \bfseries{-32:33:52.2}  & \bfseries{171.4   $\pm$  11.6} & \bfseries{52.1  $\pm$  11.2} &  \bfseries{0.30 $\pm$   0.02} &  \bfseries{222}\\
8a & 23:57:59.7 & -32:33:49.3 &     5.4 $\pm$    0.78 &     5.8 $\pm$    0.85 &   1.1 $\pm$  0.2 \\
8b & 23:57:59.8 & -32:33:54.5 &     5.6 $\pm$    0.46 &     5.1 $\pm$    0.41 &  0.92 $\pm$  0.1 \\
8c & 23:57:59.4 & -32:33:56.7 &     5.2 $\pm$    0.59 &     5.4 $\pm$    0.59 &     1 $\pm$  0.2 \\
8d & 23:57:59.0 & -32:33:52.9 &     2.6 $\pm$    0.43 &     2.8 $\pm$    0.41 &   1.1 $\pm$  0.2 \\
\hline

\multicolumn{6}{c}{NGC 55}\\
\hline
\bfseries{0} & \bfseries{00:15:47.3} & \bfseries{-39:16:28.1} & \bfseries{987.1 $\pm$   40.4} & \bfseries{507.0 $\pm$   49.3} & \bfseries{0.51 $\pm$   0.02} & \bfseries{124} \\
0a & 00:15:54.6 & -39:15:10.2 &   6.19 $\pm$   0.67 &   4.21 $\pm$   0.57 &  0.68 $\pm$  0.1 \\
0b & 00:15:54.0 & -39:15:19.7 &   19.3 $\pm$    1.1 &   13.7 $\pm$   0.96 &  0.71 $\pm$ 0.06 \\
0c & 00:15:54.5 & -39:15:29.8 &   4.18 $\pm$   0.59 &   3.49 $\pm$   0.49 &  0.83 $\pm$  0.2 \\
\hline
\bfseries{1} & \bfseries{00:15:54.9} & \bfseries{-39:15:21.6} & \bfseries{368.5 $\pm$   11.6} & \bfseries{147.9 $\pm$   13.0} &  \bfseries{0.40 $\pm$   0.01} & \bfseries{257} \\
1a & 00:15:46.8 & -39:16:28.6 &   5.39 $\pm$   0.83 &   5.87 $\pm$   0.72 &   1.1 $\pm$  0.2 \\
\hline
\bfseries{2} & \bfseries{00:15:42.3} & \bfseries{-39:16:19.5} & \bfseries{ 5.1 $\pm$   1.2} &  \bfseries{4.7 $\pm$   1.4} & \bfseries{0.91 $\pm$   0.21} &  \bfseries{41} \\
2a & 00:15:42.5 & -39:16:19.2 &    2.1 $\pm$   0.33 &   2.48 $\pm$   0.41 &   1.2 $\pm$  0.3 \\
2b & 00:15:42.3 & -39:16:18.9 &   1.88 $\pm$   0.29 &   1.95 $\pm$   0.39 &     1 $\pm$  0.2 \\
\hline
\bfseries{3} & \bfseries{00:15:39.8} & \bfseries{-39:15:21.4} & \bfseries{269.5 $\pm$   7.3} &  \bfseries{65.2 $\pm$   6.3} &  \bfseries{0.24 $\pm$   0.01} & \bfseries{97} \\
3a & 00:15:41.0 & -39:15:38.8 &   10.1 $\pm$    1.5 &   7.97 $\pm$   0.85 &  0.79 $\pm$  0.1 \\
\hline
\bfseries{4} & \bfseries{00:15:01.9} & \bfseries{-39:13:17.7} &  \bfseries{22.3 $\pm$   11.9} &  \bfseries{26.2 $\pm$   13.0} & \bfseries{1.18 $\pm$   0.74} &  \bfseries{89} \\
4a & 00:15:39.6 & -39:15:18.6 &   3.63 $\pm$   0.47 &   2.79 $\pm$   0.51 &  0.77 $\pm$  0.2 \\
\hline
\bfseries{5} & \bfseries{00:14:53.9} & \bfseries{-39:10:56.6} & \bfseries{145.0 $\pm$   14.7} &  \bfseries{63.5 $\pm$   17.2} & \bfseries{0.44 $\pm$   0.04} & \bfseries{126} \\
5a & 00:15:01.9 & -39:13:13.3 &   2.87 $\pm$   0.54 &   2.46 $\pm$   0.47 &  0.86 $\pm$  0.2 \\
\hline
\bfseries{6} & \bfseries{00:14:35.7} & \bfseries{-39:10:02.2} & \bfseries{179.7 $\pm$   25.8} & \bfseries{134.8 $\pm$   34.1} & \bfseries{0.75 $\pm$   0.11} &  \bfseries{81} \\
6a & 00:14:53.3 & -39:10:55.7 &   1.91 $\pm$   0.59 &   2.97 $\pm$   0.48 &   1.6 $\pm$  0.5 \\
\hline
\bfseries{7} & \bfseries{00:14:24.4} & \bfseries{-39:10:28.3} & \bfseries{112.1 $\pm$   10.1} &  \bfseries{86.3 $\pm$   12.8} & \bfseries{0.77 $\pm$   0.07} &  \bfseries{52} \\
7a & 00:14:35.6 & -39:09:59.5 &   5.23 $\pm$   0.48 &   3.31 $\pm$   0.49 &  0.63 $\pm$  0.1 \\
\hline
\bfseries{8} & \bfseries{00:15:41.0} & \bfseries{-39:15:38.5} &  \bfseries{26.2 $\pm$   1.4} &  \bfseries{15.0 $\pm$   1.6} & \bfseries{0.57 $\pm$   0.03} &  \bfseries{79} \\
8a & 00:14:24.2 & -39:10:28.6 &   3.86 $\pm$   0.35 &    2.50 $\pm$   0.38 &  0.65 $\pm$  0.1 \\
\hline
\multicolumn{6}{c}{NGC 1313}\\
\hline
\bfseries{0} & \bfseries{03:18:16.8} & \bfseries{-66:34:38.4} & \bfseries{153.0 $\pm$   2.8} & \bfseries{117.3 $\pm$   2.8} & \bfseries{0.77 $\pm$   0.01} & \bfseries{157} \\
0a & 03:18:17.1 & -66:34:35.2 &   5.49 $\pm$   0.82 &    4.80 $\pm$   0.62 &  0.88 $\pm$  0.2 \\
\label{table:SNRs_total_rings}
\end{longtable}
\footnotesize{The larger structures (indicated in bold), along with the respective knots that we detected. The last column gives the physical size of the structures.}
\end{ThreePartTable}

\twocolumn

\subsection{Luminosity functions (LFs)} \label{LF}
\subsubsection{H$\rm{\alpha}$ luminosity function} \label{LF_Ha}
In order to measure the H${\rm \alpha}$ luminosity function ($\rm H{\alpha} - LF$) free of selection effects, we need to apply the incompleteness function. The incompleteness-corrected number of sources is calculated by defining the quantity 1/$\rm I_i$ where $\rm I_i$ is the completeness corresponding to each source, given their  [\ion{S}{II}], H${\rm \alpha}$ luminosity (see $\S$ \ref{incomp}). The completeness of each source  depends on its local background. As described in $\S$ \ref{incomp}, we derive completeness maps for four different background regimes. 

\par
\begin{figure*}

            \includegraphics[width=0.475\textwidth]{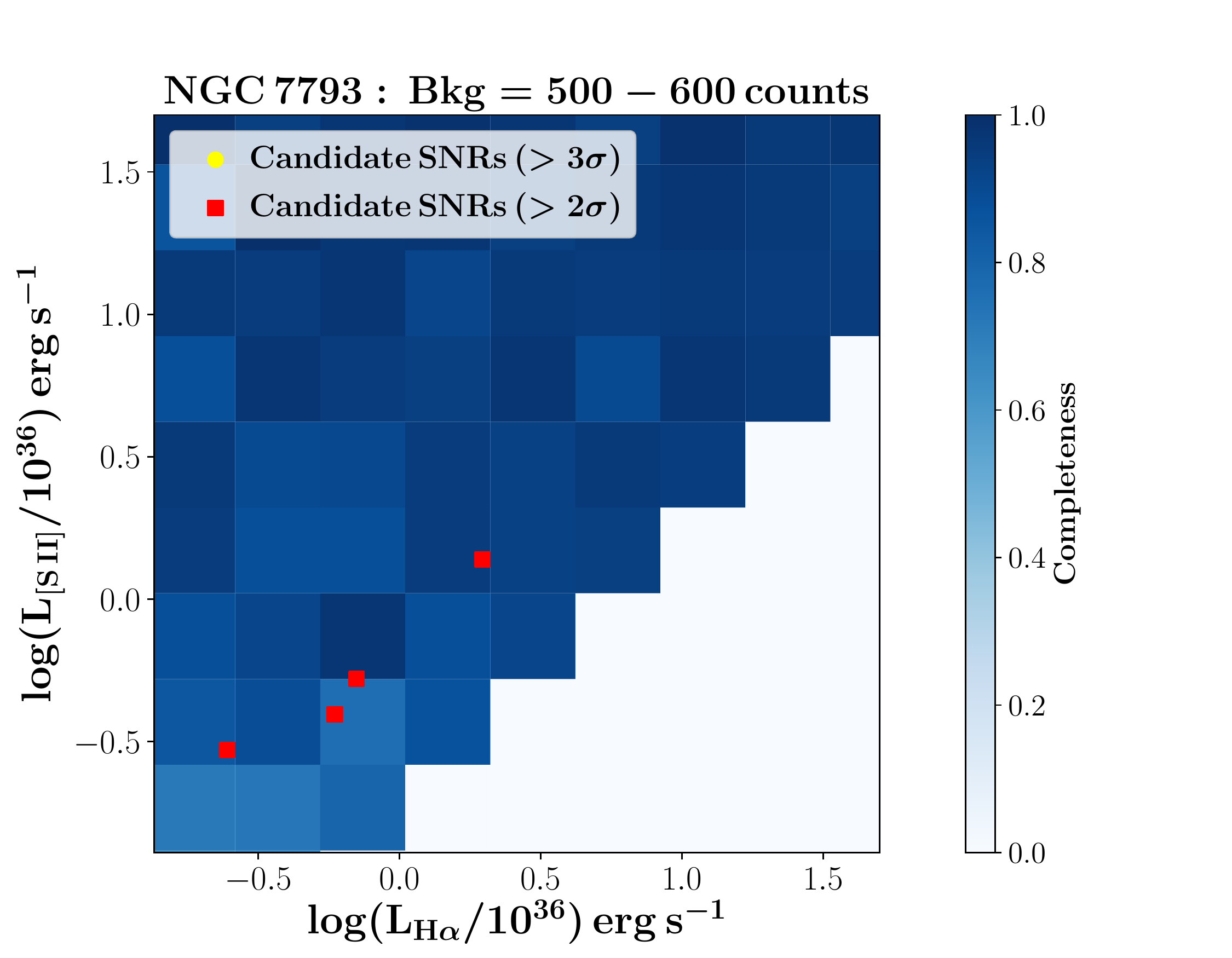}%
            \includegraphics[width=0.475\textwidth]{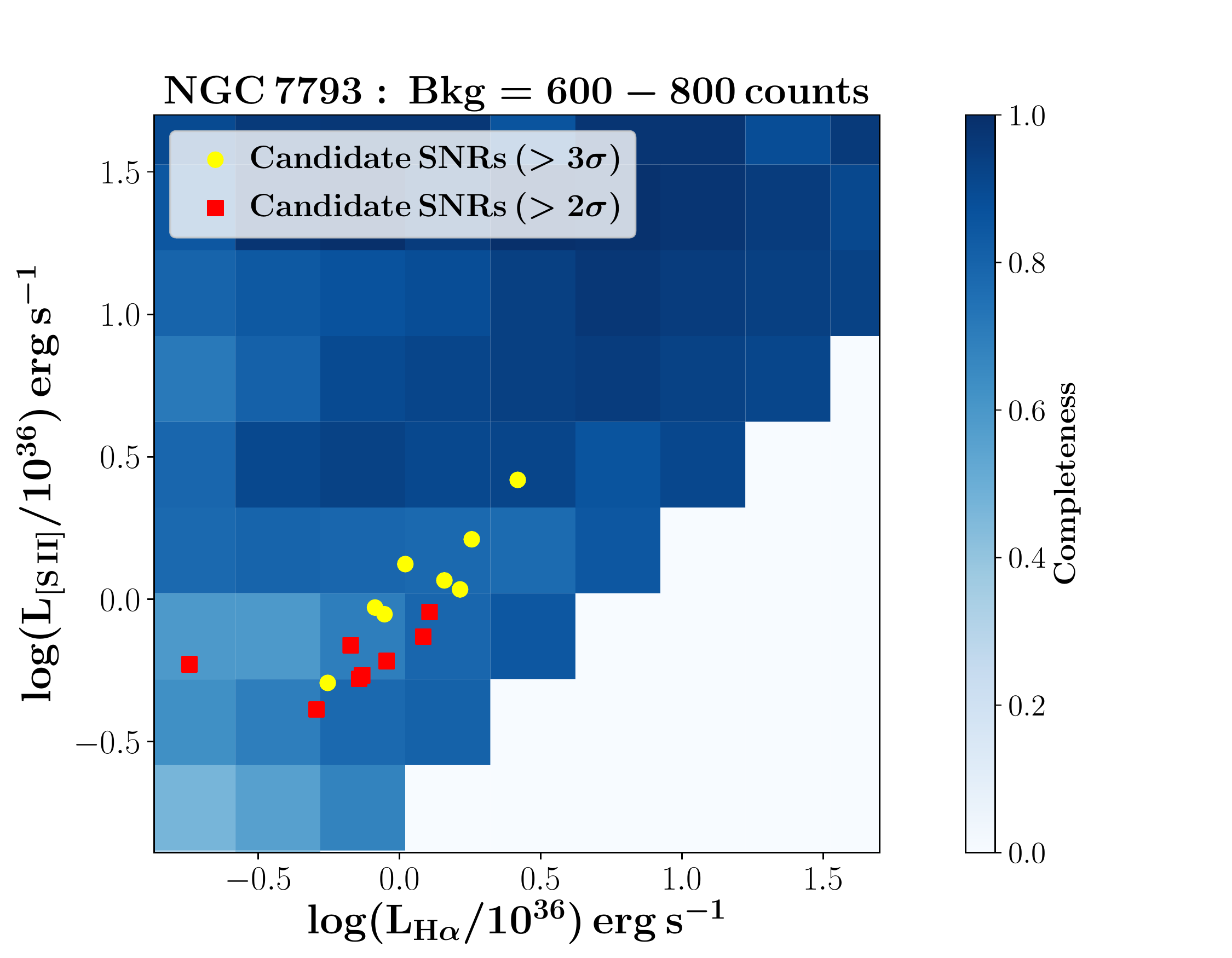}\hfill

            \includegraphics[width=0.475\textwidth]{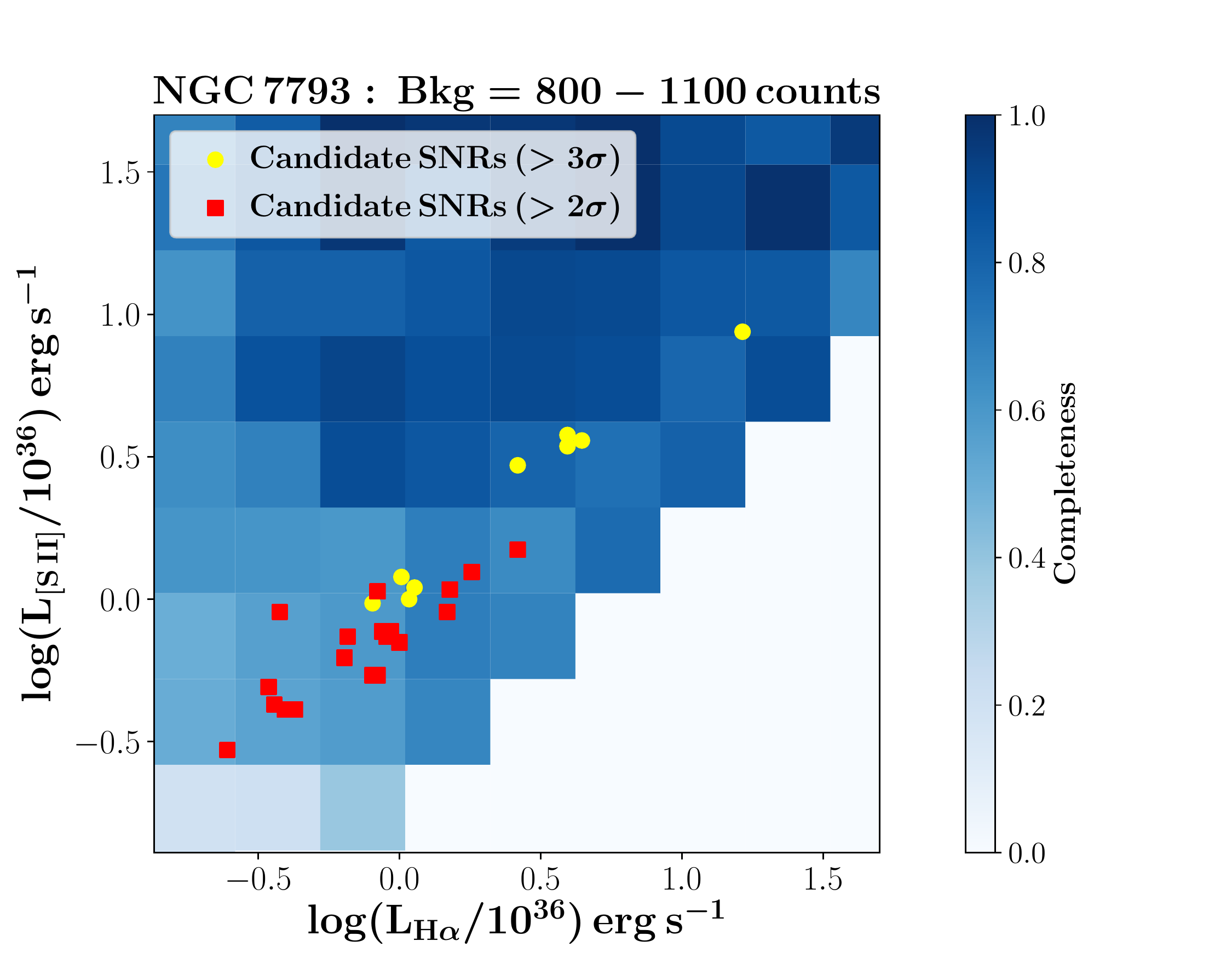}
            \includegraphics[width=0.475\textwidth]{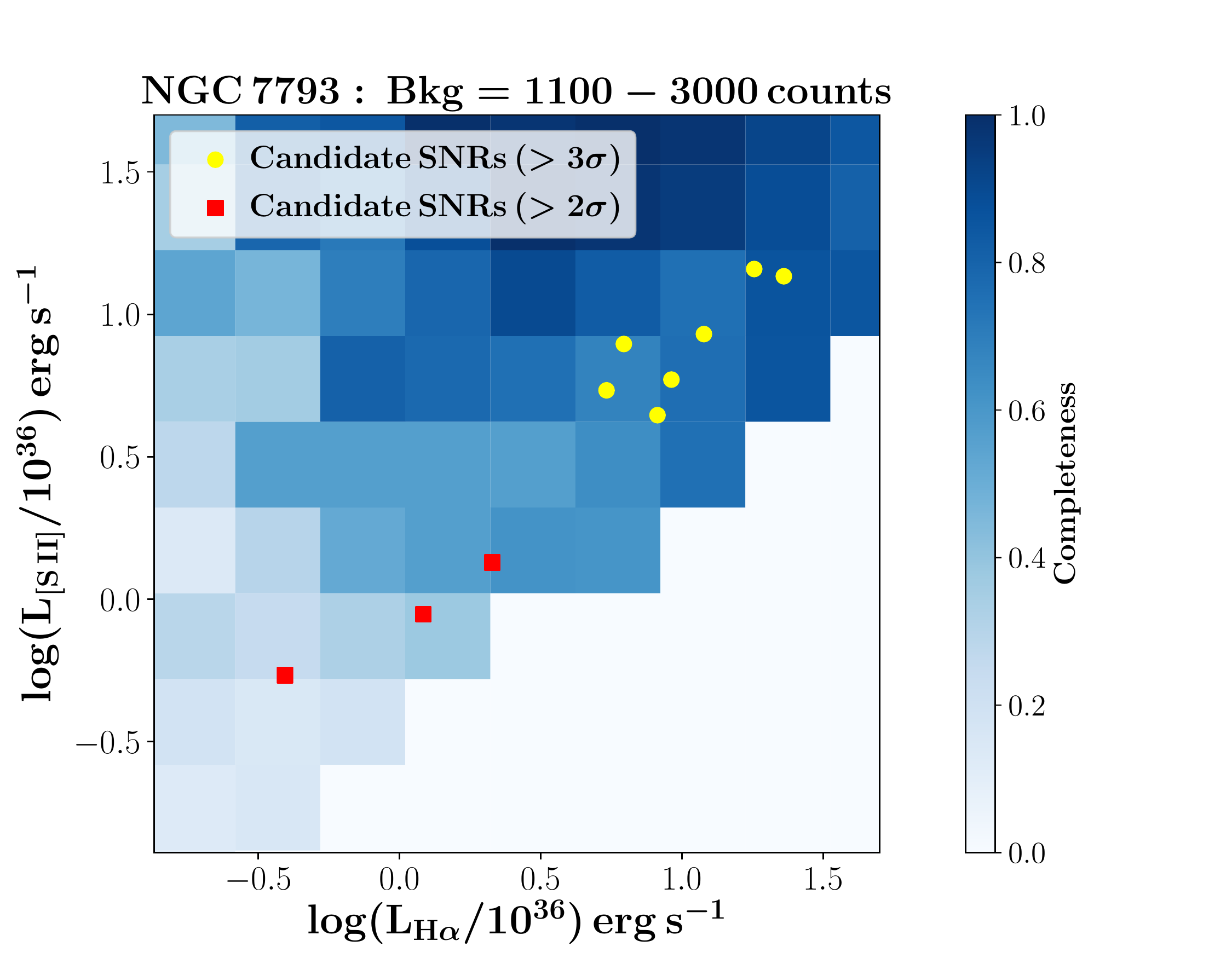}

        \caption[]
        {Two dimensional H$\rm \alpha$ - [\ion{S}{II}] incompleteness maps for the candidate SNRs (with [\ion{S}{II}]/H$\rm \alpha$ ratio 3$\rm \sigma$ and 2$\rm \sigma$ above 0.4 respectively) in NGC 7793, for 4 different backgrounds. Darker colors indicate lower incompleteness (more complete samples). The points show the location of the candidate SNRs (yellow circles) and possible candidate SNRs (red squares) on the (H$\rm \alpha$, [\ion{S}{II}]) plane.}     
        \label{fig:incompleteness}
    \end{figure*}

 Indicatively, in \autoref{fig:incompleteness} we present  the incompleteness maps for NGC 7793 for the different backgrounds. The x and y axis correspond to the logarithm of the H$\rm{\alpha}$ and [\ion{S}{II}] luminosities respectively, and the degradation of the blue scale indicates the completeness for each combination of [\ion{S}{II}] and H$\rm \alpha$ luminosity. On these maps, we also show the detected candidate SNRs (> 3$\rm \sigma$; yellow circles) and the possible candidate SNRs (> 2$\rm \sigma$; red squares). As we see, the highest completeness is close to 100\%. 

\par For the calculation of H$\rm \alpha$ LF we consider only the candidate SNRs (and not the possible candidate SNRs), and we also exclude any sources associated with larger structures (\autoref{table:SNRs_total_rings}) since they could be knots in larger filaments or super-bubbles. The incompleteness maps that we use  to correct these sources, have been constructed using artificial objects that satisfy exactly the same criteria used to select the candidate SNRs (i.e. [\ion{S}{II}]/H${\rm \alpha}$ ratio, 3$\rm \sigma$ above 0.4).  The larger structures are not included in this analysis, since our detection method is not optimized for extended objects and we cannot apply the incompleteness correction, which is based on point-like  sources (\S    \ref{incomp}).

\par
\begin{figure}
            \includegraphics[width=0.45\textwidth]{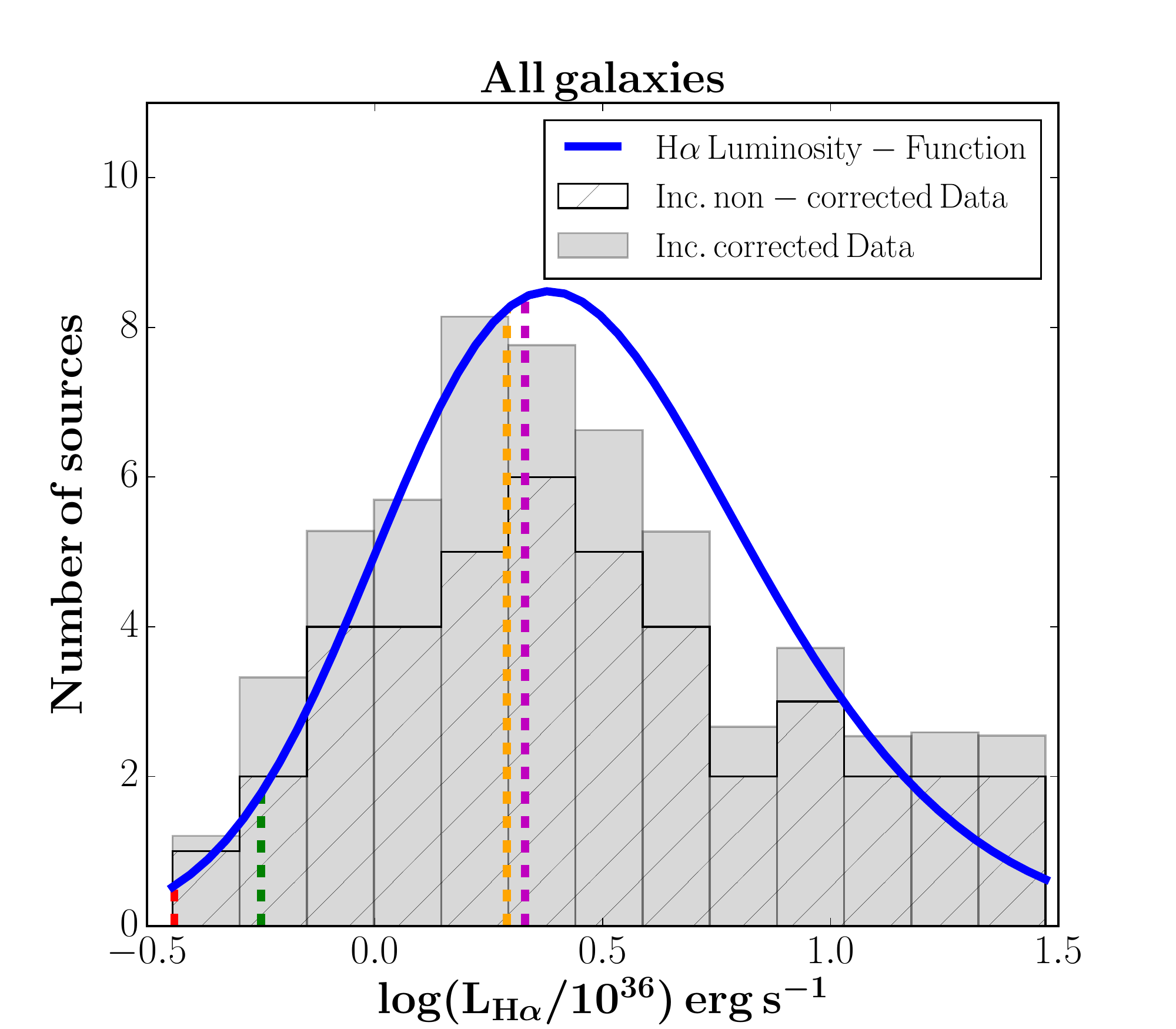}
        \caption[]
        {The incompleteness-corrected H$\rm \alpha$ luminosity function of the candidate SNRs (with ([\ion{S}{II}]/H$\rm \alpha$ - 0.4) > 3$\rm \sigma$; gray histogram), plotted along with the observed luminosity function without any incompleteness correction (black-line histogram).  The blue curve shows the best-fit $\rm LF_{H\alpha}$ fitted to the incompleteness-corrected data. The vertical dashed-lines present the faintest H$\rm \alpha$ luminosity limits of the candidate SNRs in the galaxies NGC 55 (red), NGC 7793 (green), NGC 1313 (orange), and NGC 45 (magenta).}
        \label{fig:lf_Ha-both}
    \end{figure}

 \autoref{fig:lf_Ha-both} shows the binned differential incompleteness-corrected H${\rm \alpha}$ luminosity functions for the candidate SNRs in all galaxies (gray histogram).  The incompleteness corrected number of sources for each luminosity bin is given by the sum $\rm \sum _i (1/I_i)$ for all sources within each H$\alpha$ bin. For reference, the solid-line histogram shows the observed (uncorrected) binned H${\rm \alpha}$ luminosity distribution. We see, that as expected, in general, the effect of incompleteness is more significant for the lower luminosity sources. However, even in the bright end the completeness is never 100\% because of the selection in the [\ion{S}{II}] emission lines and the high background in which we find the more luminous SNRs. The vertical dashed-lines show the H$\rm \alpha$ luminosity limit for the different galaxies (red, green, orange, and magnenta for NGC 55, NGC 7793, NGC 1313, and NGC 45 respectively). The fact that in the H$\rm \alpha$ LF we include also the galaxies  NGC 45 and NGC 1313 for which the lowest luminosity limit is higher than  NGC 55 and NGC 7793, does not affect significantly the shape of the H$\rm \alpha$ LF, since these galaxies  have a very small number of SNRs. Furthermore the LF does not show any  features at the location of the luminosity limits for each sample. In fact jointly fitting the four individual LFs (leaving free their normalizations) gives very similar results for their shape parameters.

\par We fit the $\rm  LF_{H{\alpha}}$ by performing a maximum likelihood fit on the un-binned data. We fit a model described by a skewed Gaussian function (in the log(L)) in order to account for the asymmetric shape of the LFs (\autoref{fig:lf_Ha-both}). The skewed-Gaussian distribution is given by the function:
\begin{equation} 
\rm{f(logL) = \frac{A}{\sigma \sqrt{(2\pi)}}e^{\frac{-(logL-\mu)^2}{2\sigma ^2}}\{1 + erf[\frac{\alpha (logL - \mu)}{\sigma \sqrt{2}}]\}} \label{eq:1}
\end{equation}
where  $\rm \mu$ is the mean of the distribution, $\rm \sigma$ the standard deviation, and $\rm \alpha$ the skewness parameter which describes the asymmetry of the LF and potential selection effects. The incompleteness is included as a weight term in the likelihood function.

\par The amplitude (A), of the function reflects the total number of SNRs ($\rm N_{Total}$). 
The number of incompleteness corrected objects down to our faintest limit is calculated by $\rm K = \sum _{i=0} ^N (1/I_i)$, where N is the total number of observed sources in each galaxy. Since the function used to model the luminosity function is normalized, the total number of incompleteness corrected objects (even beyond the faintest limit) and hence the amplitude factor A in \cref{eq:1} is given by $\rm A = N_{Total} = K/\int_{log{(L_{min})}}^{log{(L_{max})}} f(log(L)) d(log(L))$, where $\rm L_{min}$ and $\rm L_{max}$ are the minimum and maximum luminosity (in this case H$\alpha$ luminosity) of our sample.

\par For the determination of the best fit model parameters we used a Markov Chain Monte Carlo method implemented in  the python package \texttt{pymc}\footnote{https://pymc-devs.github.io/pymc/modelfitting.html}. We assume flat priors for all the parameters of interest ($\rm \mu,\, \sigma, \, and \, \alpha$) and we perform 40000 iterations (excluding 500 burn-in iterations). The best-fit parameters and their uncertainties (at the 68\% percentile) as derived by the posterior distribution of the model parameters are given in \autoref{table:LHa_par}. The best-fit model to the un-binned data is shown as the solid line in \autoref{fig:lf_Ha-both}.

\par  In \autoref{table:LHa_par} we also give the number of the detected SNRs (N) and the total number of SNRs (A; i.e. after accounting for incompleteness) of the overall sample of the 4 galaxies, and we calculate the detection fraction, which is the ratio of the number of the detected sources over the total number of SNRs ($\rm N/A$). As we see, the detected SNRs can be $\sim$ 30\% less than the total number of SNRs in a galaxy. We can also estimate the total, incompleteness corrected, number of SNRs for each galaxy individually, down to the lowest H$\rm \alpha $ luminosity limit of the global H$\rm \alpha $ LF. To do that, we calculate the $\rm{N_{i}}\times \frac{\int_{L_{min,i}}^{\infty}LF(L_{H\rm \alpha})dL_{H\rm \alpha}}{\int_{L_{min}}^{\infty}LF(L_{H\rm \alpha})dL_{H\rm \alpha}}$,
where ${L_{min,i}}$ is the lowest H$\rm\alpha$ luminosity of each galaxy $i$,  $N_i$ is the incompleteness corrected number of sources of each galaxy, and ${L_{min}}$ is the lowest H$\rm\alpha$ luminosity of our total sample.
This analysis gives 14.8, 5.8, 13.5, and 32.6 candidate SNRs for NGC 45, NGC 55, NGC 1313, and NGC 7793 respectively.

\par If we take into account also the possible candidate SNRs ([\ion{S}{II}]/H$\rm\alpha$ ratio 2$\rm \sigma$  above the 0.4 threshold) we find for the H$\rm\alpha$ LF $\rm \mu = -0.37^{+0.79}_{-0.01},\, \sigma = 0.61^{+0.17}_{-0.03}$ and $\rm \alpha = 1.09^{+0.20}_{-0.50}$.


\begin{table*}
\begin{threeparttable}
\centering
\caption{H$\rm{\alpha}$ Luminosity Function parameters for candidate SNRs (>3$\rm\sigma$)}
\begin{tabular}{lccccccc}
\hline
 N & $\rm{\mu}\,(log(L^{*}))$ & $\rm{\sigma}\,(log(L^{*}))$ & $\rm{\alpha}$ & $\rm N_{max} $ & $\rm A $ & Detection Fraction\\ 
\hline

 42 &	 $\rm 0.07 ^{+0.63} _{-0.05}$ & $\rm 0.58 ^{+0.11} _{-0.09}$ &	 $\rm 1.70 (>0.48)$ &	 $\rm  8.62 \pm 1.32$ & 59.83  $\pm$ 9.17 & 0.70  $\pm$ 0.10\\
\hline
\end{tabular}
\label{table:LHa_par}
\begin{tablenotes}
\item Col (1): the number of candidate SNRs (> 3$\rm \sigma$) of all galaxies; col (2) and col (3): the mean and sigma of the H$\rm \alpha$ luminosity function respectively; col (4): the skewness parameter of the H$\rm \alpha$ luminosity function; col (5): the model-based number of sources at the bin corresponding to the peak of the distribution; col (6): the total number of SNRs; col (7): the ratio of the number of the detected sources over the total number of SNRs  ($\rm N/A$).\\
 $^{*}$Luminosity in units of $\rm 10^{36}$ \ergs\\
 \end{tablenotes}
 \end{threeparttable}
\end{table*}

\subsubsection{The joint [S II] - H$\rm{\alpha}$ luminosity function} \label{SII - Ha luminosity function}
Since we have measurements of both the H$\rm \alpha$ and [\ion{S}{II}] luminosities of the SNRs in our sample, we can also calculate their joint luminosity function (LF) which yields information on the luminosity and excitation of the SNR populations.  The location of the sources on the H$\rm \alpha$ - [\ion{S}{II}] luminosity plane is shown in \autoref{fig:lines}.

\par  To quantify the shape of the [\ion{S}{II}]-H$\rm \alpha$ joint distribution for the candidate SNRs (and not the possible candidate SNRs), we adopt the following approach. 
We first calculate the relation between the $\rm log(L_{[S\, II]})$ and $\rm log(L_{H\alpha})$ luminosity for the detected sources. The best-fit relation is log([\ion{S}{II}]) = 0.88log(H$\rm \alpha$) - 0.06. This is the "spine" of the distribution of the candidate SNRs on the [\ion{S}{II}]-H$\rm \alpha$ plane, shown by the blue line in \autoref{fig:lines}.

Then, we project all the points on the best-fit line (i.e. for each H$\rm \alpha$ we calculate the log([\ion{S}{II}]) - y = log([\ion{S}{II}]) - (0.88log(H$\rm \alpha$) - 0.06)).  We then fit a skewed Gaussian to the distribution of these points along the best-fit line, applying the incompleteness correction for each source (considering each source $\rm i$ as $\rm 1/Ii$; $\S$ \ref{LF_Ha}). As for the H$\rm \alpha$ LF, we use a Markov Chain Monte Carlo method, assuming flat priors for all the parameters of interest ($\rm \mu,\, \sigma, \, and \, \alpha$) and performing 40000 iterations (excluding 500 burn-in iterations). The best-fit parameters of the 2D LF, are presented in \autoref{table:LF_par}. If we take into account also the possible candidate SNRs ([\ion{S}{II}]/H$\rm\alpha$ ratio 2$\rm \sigma$ above the 0.4 threshold) the the parameters $\rm \mu,\, \sigma$ and $\rm \alpha$ parameters of the 2D [\ion{S}{II}] - H$\rm\alpha$ LF are $-0.34^{+0.66}_{-0.04}$, $0.53^{+0.14}_{-0.03}$ and $0.57^{+0.67}_{-0.07}$ respectively.

\par Aiming to obtain a full picture of the 2D H$\rm \alpha$ - [\ion{S}{II}], we also need to measure the scatter of the candidate SNRs around the "spine" of the [\ion{S}{II}] - H$\rm \alpha$ distribution (the best-fit line derived above). In order to do so, we calculate the distance of each source from the  best-fit [\ion{S}{II}] - H$\rm \alpha$  line, on the [\ion{S}{II}] axis. The distribution of these distances is described by a truncated Gaussian the fit parameters of which have been calculated using the same approach as above. The $\rm \mu$ and $\rm \sigma$ values of the truncated Gaussian are $0.024^{+0.025}_{-0.024}$ and $0.14^{+0.018}_{-0.029}$ respectively. The sigma  is considered as the width of the 2D LF.  The joint [\ion{S}{II}]-H$\rm \alpha$ LF is presented in \autoref{fig:LF_3d_all}. 


\par The total number of SNRs may be affected by the number of false positives: i.e. sources which satisfy the [\ion{S}{II}]/H$\rm \alpha$ > 0.4 criterion even at the 3$\rm \sigma$ level, while in reality they have lower [\ion{S}{II}]/H$\rm \alpha$ ratios. Such sources could be \ion{H}{II} regions, with [\ion{N}{II}]/H$\rm \alpha$ ratio different than the one that we have adopted for our analysis. However, based on the analysis presented in \S \ref{incomp} we calculate that the fraction of sources with intrinsic [\ion{S}{II}]/H$\rm \alpha$ ratios between 0.2-0.3 and 0.3-0.4, that are found to satisfy our [\ion{S}{II}]/H$\rm \alpha$ > 0.4 criterion at the 3$\rm \sigma$ level, is only 0.3\% - 1.0\%  for the different galaxies in our sample and for the different backgrounds in each galaxy. Therefore, we consider the contamination by lower excitation sources negligible.

\begin{table*}
\begin{threeparttable}
\centering
\caption{[\ion{S}{II}] - H$\rm{\alpha}$ Luminosity Function parameters for candidate SNRs (>3$\rm\sigma$)}
\begin{tabular}{clccccccc}
\hline
 $\rm Axis$ & $\rm N $ & $\rm{\mu}\,(log(L^{*}))$ & $\rm{\sigma}\,(log(L^{*}))$ & $\rm \alpha$ & $\rm{N_{max}}$  & $\rm A $ & Detection fraction\\ 
\hline
$\rm L = 0.88log(L_{H\alpha}) - 0.06$ & 42 & $-0.03^{+0.56}_{-0.03}$  & $0.48^{+0.13}_{-0.05}$ & 1.73(>1.69) & 7.62 $\pm$ 1.17 & 59.38 $\pm$ 9.10 & 0.71    $\pm$ 0.01\\
\\
$\rm log(L_{[S\, II]}/10^{36})$ & - & $0.024^{+0.025}_{-0.024}$  & $0.14^{+0.02}_{-0.03}$ & - & - & - & - \\
\hline
\end{tabular}
\label{table:LF_par}
\begin{tablenotes}
\item Col (1): the number of candidate SNRs (>3$\rm\sigma$) of all galaxies; col (2) and col (3): the mean and sigma values of the joint [\ion{S}{II}] - H$\rm \alpha$ luminosity function respectively (for the skewed Gaussian along the line $\rm y = 0.88x - 0.06$ which shows the shape of the LF; first row, and for the truncated Gaussian on the [\ion{S}{II}] axes, i.e the distance of the  [\ion{S}{II}] from the line $\rm y = 0.88x - 0.06$, which shows the width of the LF; second row); col (4): the skewness parameter of the joint [\ion{S}{II}] - H$\rm \alpha$ luminosity function; col (5): the model-number of the sources at the bin that corresponds to the peak of the distribution; col (6): the total number of candidate SNRs (even beyond the lower limit); col (7): the ratio of the number of the detected sources over the total number of candidate SNRs  ($\rm N/A$).\\
$^{*}$ $\rm L = 0.88log(L_{H\alpha}) - 0.06$ (first row), and $\rm L = L_{[S\,II]}$ (second row),  in units of $\rm 10^{36}$ \ergs
 \end{tablenotes}
\end{threeparttable}
\end{table*}

\subsection{Excitation Function} \label{excitation}

\par An alternative way to view the 2D luminosity function, is in terms of the excitation of the SNRs, which is described by their [\ion{S}{II}]/ H$\rm \alpha$ ratio. The ratio of the forbidden and Balmer lines is an indicator of the degree of excitation of the emitting material. Therefore the range of this ratio can be used as a proxy for the physical condition in the SNR population (particularly in combination with shock excitation models). For this reason, we calculate the number density of SNRs, as function of their distance from the limiting [\ion{S}{II}]/H$\rm \alpha$ = 0.4 line, which is the threshold for SNR classification. 

\begin{figure*}

\begin{minipage}{0.48\textwidth}

\includegraphics[width=0.95\textwidth]{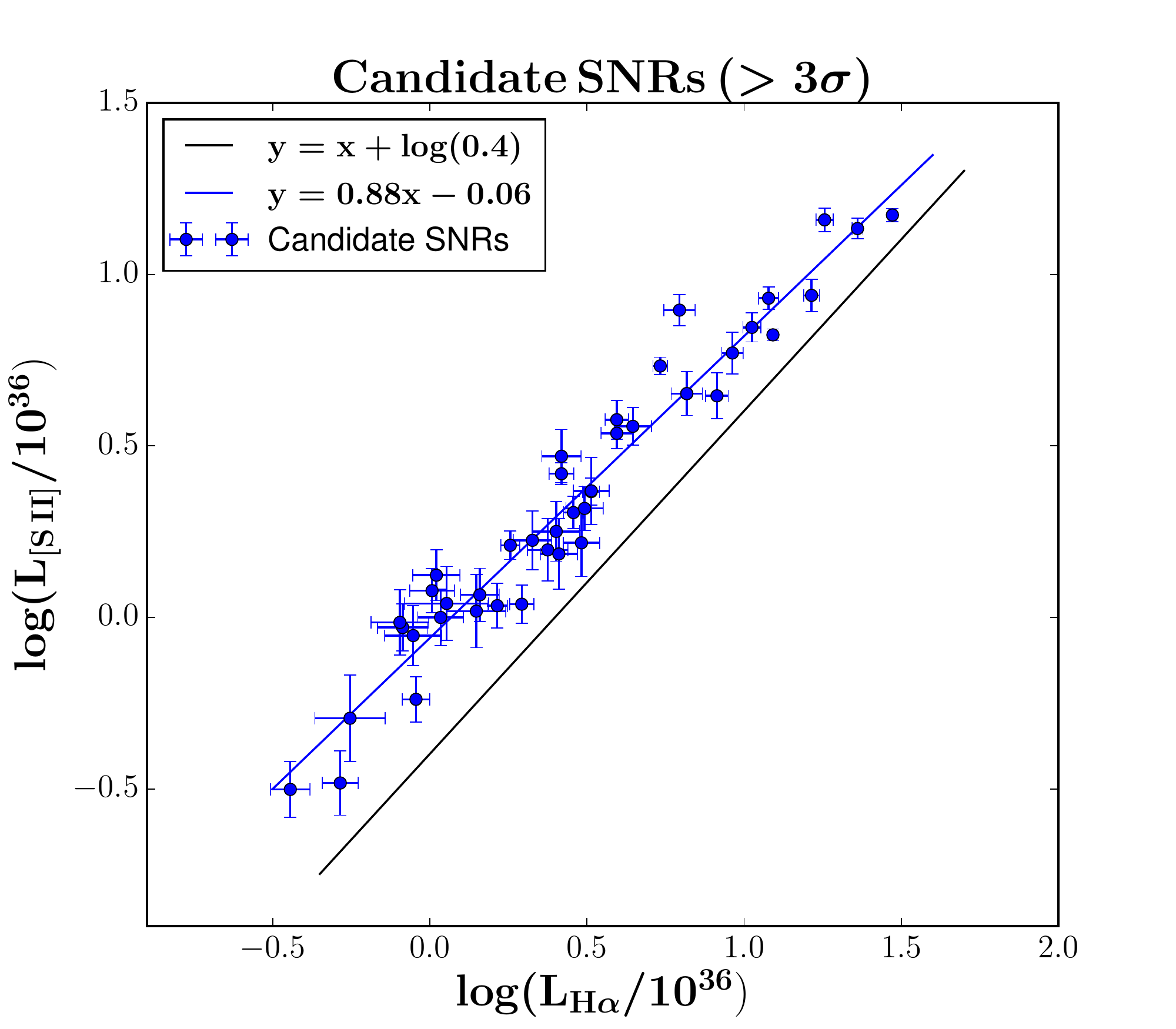}
 \caption{{A scatter plot of the [\ion{S}{II}] and H$\rm\alpha$ luminosity of the candidate SNRs.  The black line shows the [\ion{S}{II}]/H$\rm\alpha$ = 0.4 threshold. The blue line shows the best-fit log($\rm L_{[S\, II]}$) - log($\rm L_{[H\alpha}$) relation for the sample of candidate SNRs.}}
        \label{fig:lines}
\end{minipage}\hfill
\begin{minipage}{0.48\textwidth}
\includegraphics[width=1\textwidth]{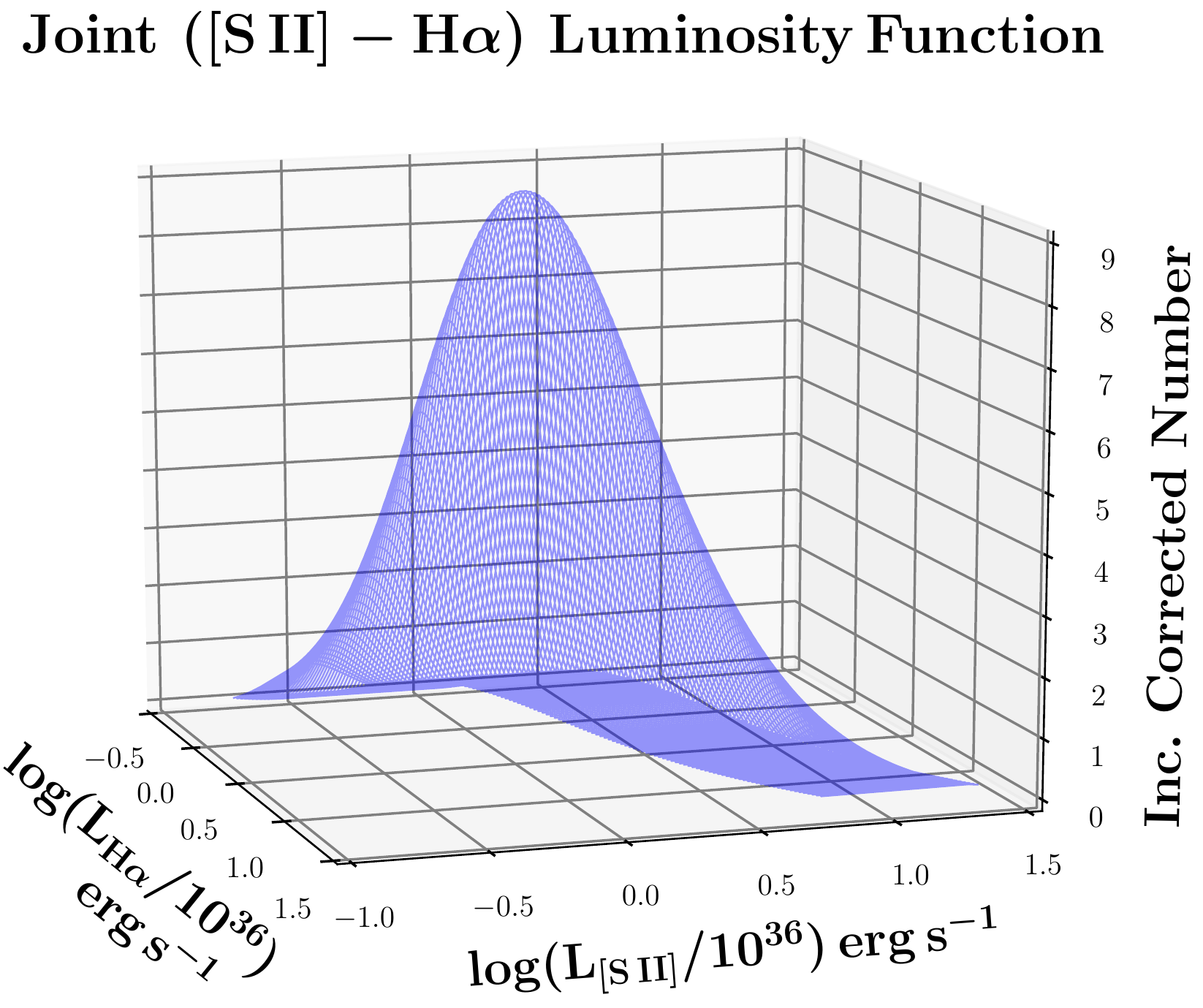}
\caption[]
        {{The joint [\ion{S}{II}] - H$\rm \alpha$ luminosity function along the best-fit line: $\rm log(L_{[S\, II]}/10^{36}) = 0.88log(L_{[H\alpha}/10^{36})-0.06$. The parameters of the joint LF are presented in \autoref{table:LF_par}. In order to obtain the 3D interpretation, we multiply the skewed Gaussian that describes the shape of the joint LF along the best-fit line, with the truncated Gaussian that describes the width of the joint LF on the [\ion{S}{II}] dimension.}}
        \label{fig:LF_3d_all}
\end{minipage}

\begin{minipage}{0.48\textwidth}
\includegraphics[width=0.95\textwidth]{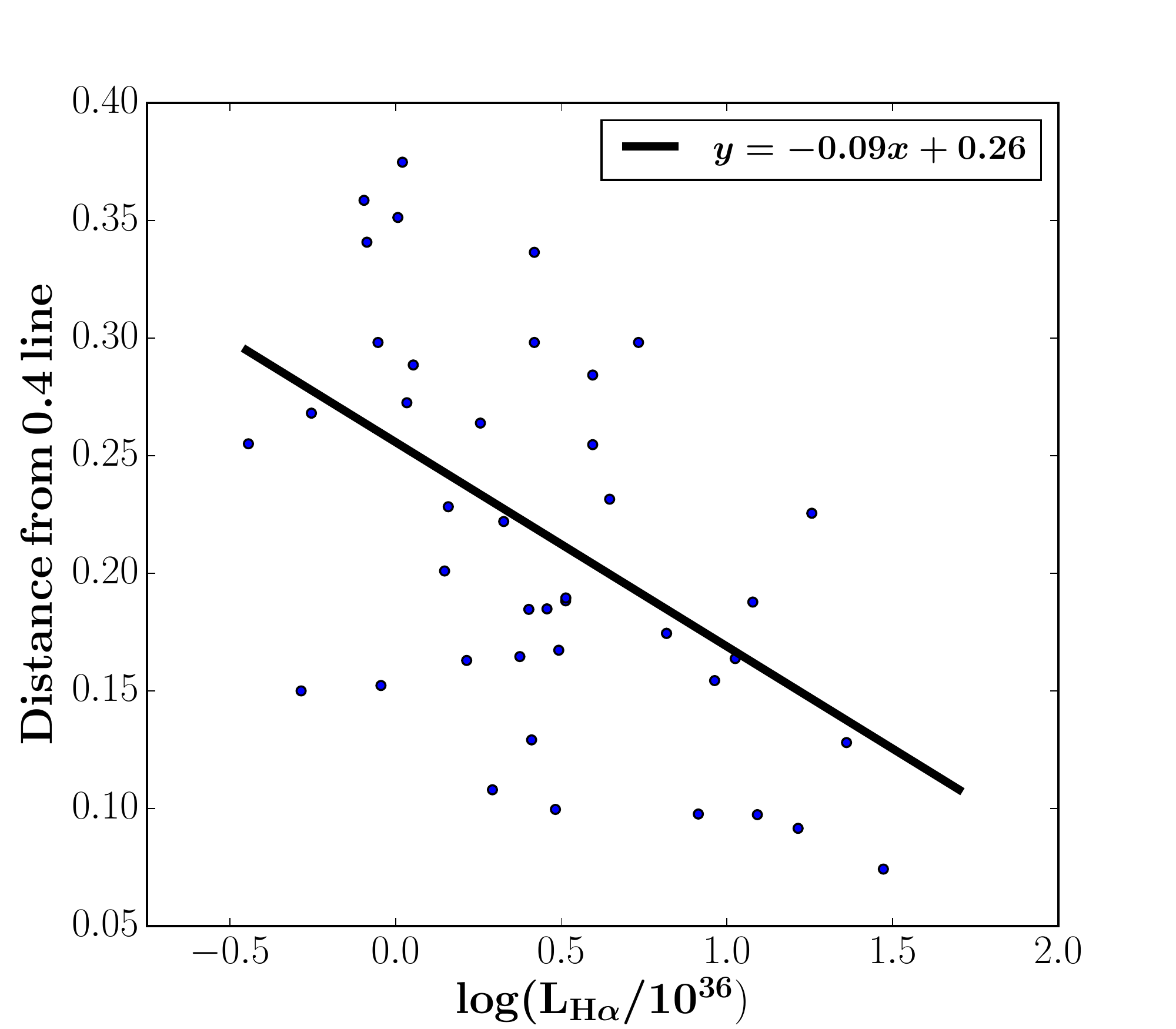}
    \caption{The points show the distance of each source from the [\ion{S}{II}]/H$\rm \alpha$ = 0.4 line. The black line is the best fit line that describes the distance of the candidate SNRs from the [\ion{S}{II}]/H$\rm \alpha$ = 0.4 line as function of their H$\rm\alpha$ luminosity.}\label{fig:Ha_dist}
\end{minipage}\hfill
\begin{minipage}{0.48\textwidth}
\includegraphics[width=0.95\textwidth]{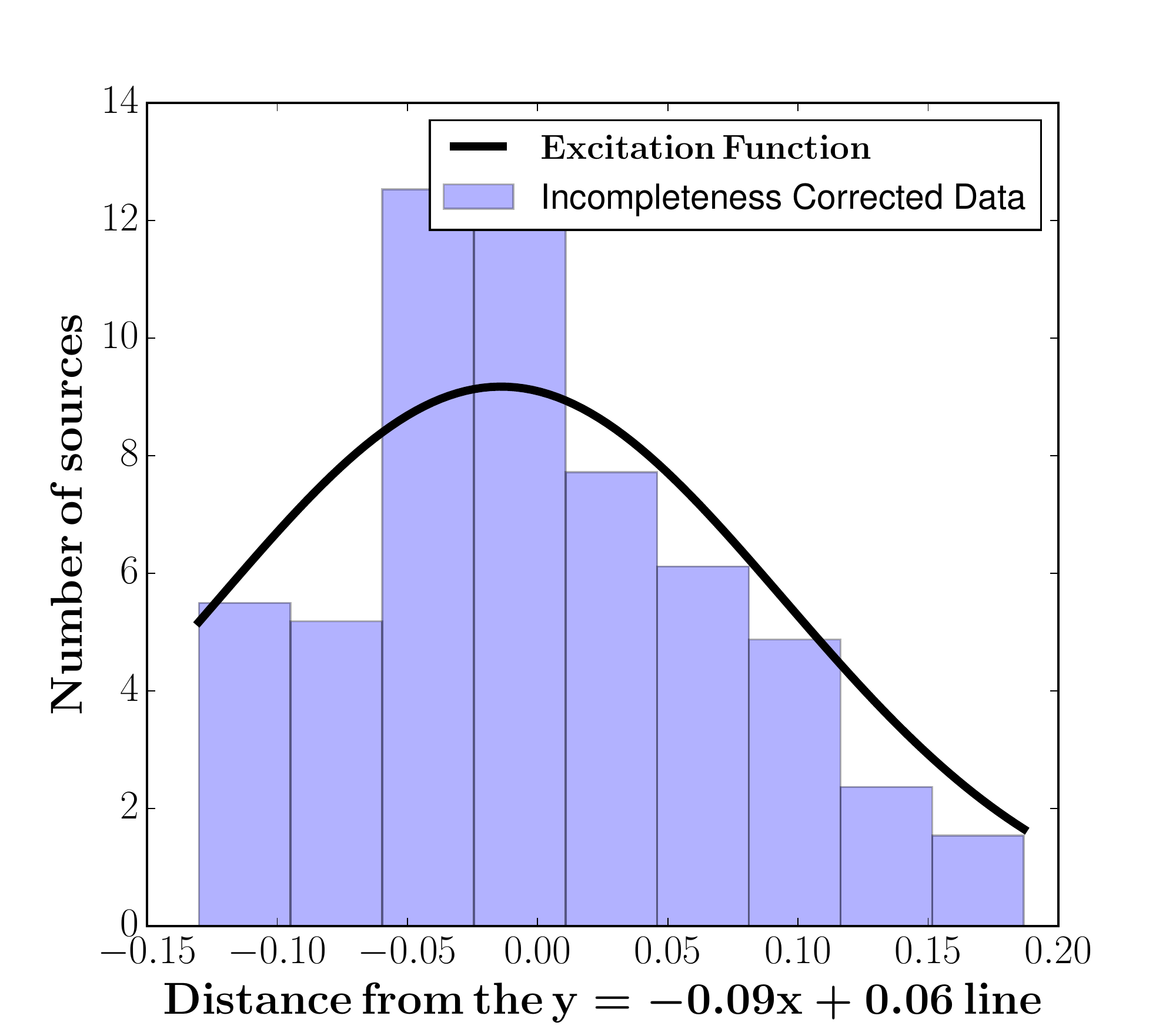}
    \caption{The histogram is the incompleteness-corrected source distance from the best-fit line of \autoref{fig:Ha_dist} distribution ($\S$ \ref{excitation}), and the black line is the fitted excitation function, described by a truncated Gaussian with $\rm \mu(log(L_{[S\, II]}/10^{36})) = 0.024$ and $\rm \sigma(log(L_{[S\, II]}/10^{36})) = 0.14$. }\label{fig:distances_dist}
\end{minipage}
\end{figure*}

In \autoref{fig:lines}, this distance corresponds to different [\ion{S}{II}]/H$\rm \alpha$ ratios (i.e. different excitation). As we can see, there is a trend for SNRs with higher H$\rm \alpha$ luminosity, to present increasingly lower excitation. To quantify this trend, we fit a linear relation to the orthogonal distances from the 0.4 line, as function of the H$\rm \alpha$ luminosity: D $=$ $\rm \alpha $log(H$\rm \alpha$) $+$ $\rm \beta$ (\autoref{fig:Ha_dist}).
We find $\rm \alpha = -0.09$ and $\rm \beta = 0.26$, indicating a sub-linear relation, i.e. more luminous objects tend to have lower excitation.

\par We introduce the excitation function as the vertical distribution of the candidate SNRs with respect to  $\rm D=-0.09log(H\alpha)+0.26$ (\autoref{fig:Ha_dist}). Then, we calculate the distance of each source from this line, in order to quantify the spread around the excitation function.   The difference between the excitation function and the 2D [\ion{S}{II}] - H$\rm \alpha$ LF, is that in the latter, the y-dimension corresponds to the [\ion{S}{II}] luminosity. 
While one can derive the excitation function from the 2D [\ion{S}{II}] - H$\rm \alpha$ LF, for simplicity and more direct assessment of the excitation of SNRs, we also introduce the excitation function.
The distribution of these distances are presented in \autoref{fig:distances_dist} (blue histogram). 

This distribution is described by a truncated Gaussian, the best parameters of which are: $\rm \mu = -0.014^{+0.011}_{-0.051}\, and\, \sigma = 0.11^{+0.04}_{-0.21}$. The fitted truncated Gaussian has been calculated for the incompleteness-corrected candidate SNRs and it is presented in \autoref{fig:distances_dist} (black line).  If we take into account also the possible candidate SNRs ([\ion{S}{II}]/H$\rm\alpha$ ratio 2$\rm \sigma$ above the 0.4 threshold) the parameters $\rm \mu$, and $\rm \sigma$ of the excitation function are $-0.004^{+0.009}_{-0.014}$, and $0.10^{+0.01}_{-0.19}$ respectively. In this case the  sub-linearity is characterized by a slope -0.10.

\section{Discussion} \label{discussion}

\subsection{Comparison with other surveys}
In our survey we have detected in total 42 candidate SNRs (30 of which are new identifications) and 54 new possible candidate SNRs. This is a significant contribution to the number of known optical extragalactic SNRs. In addition, we find 21 sources with [\ion{S}{II}]/H$\rm\alpha$ > 0.4 at both 3$\rm \sigma$ and 2$\rm \sigma$ levels, which are associated with larger structures. The size of these structures ranges from $\sim$ 40 to $\sim$ 320 pc. Larger structures with size between $\sim$40 and $\sim$100\,pc are most probably more evolved SNRs (\citealt{2012AJ....143...85F}), while those with diameter larger than 100 pc are likely super-bubbles or groups of SNRs. 
For the largest such SNRs, with diameter of $\sim$100\, pc, and assuming that the density of the surrounding ISM is 10 $\rm cm^{-3}$ and the explosion energy was $\rm 10^{51}\, erg$, according to Sedov-Taylor solution (\citealt{2011piim.book.....D}), its age is $\sim$100000 years, i.e. an old SNR.

\par In this study we present for the first time optical SNR candidates in NGC 45, NGC 55, and NGC 1313. NGC 7793 has been studied before by \citet{Blair1997} who presented 27 SNRs. Almost all these SNRs have been also identified in our deeper survey which yields 24 candidate SNRs (13 of which coincide with those of \citealt{Blair1997}) and 31 possible candidate SNRs. Five of the SNRs that are reported in the work of \citet{Blair1997} (s2, s3, s7, s18, s2, s26ext) coincide with 5 (out of 8) larger structures that we detected in this galaxy. Some SNRs in the study of \citealt{Blair1997} have [\ion{S}{II}]/H$\rm{\alpha}$ ratio less than 2$\rm{\sigma}$ above the 0.4 limit in our study so they are not discussed in this paper (s9, s17, s21, s22). We do not identify as SNRs sources  s19, s23, s15 in the catalogue of \citet{Blair1997}. Sources s4, s19 and s15 have [\ion{S}{II}]/H$\rm{\alpha}$ ratio  0.36, and 0.17 respectively, i.e. below our 0.4 limit. Source s23 instead, is resolved into multiple sources in our images none of which gives high [\ion{S}{II}]/H$\rm{\alpha}$ ratio. Source s27 in our images appears as a faint ring. Since our study focuses on more point-like sources, we do not detect shell-like sources (unless they present knots that satisfy our selection criteria as discussed in  $\S$ \ref{results}), which however, are not included in our luminosity and excitation functions.

\par Large samples of extragalactic SNRs have been presented in many surveys. In M31 (\citealt{2014ApJ...786..130L}) and M33 (\citealt{2014ApJ...793..134L}) have been detected 156 and 199 photometric SNRs respectively.  \citet{1997ApJS..112...49M} identified 93 SNRs in M101 and 41 SNRs in M81. Surveys led by \citet{2014ApJ...788...55B}, \citet{2010ApJ...710..964D} and \citet{2004ApJS..155..101B} have given 296 SNRs in M83. \citet{2013MNRAS.429..189L}, in a survey of 6 galaxies presented 149 photometric SNRs in NGC 2403, 92 in NGC 4214, 70 in NGC 4449, 47 in NGC 4395, 36 in NGC 5204 and 24 in NGC 3077. The most recent work on NGC 6946, gave 147 photometric SNRs (\citealt{Long2019}). The number of detected SNRs depends on various characteristics of a galaxy (e.g.  SFR; ISM density distribution etc.) but also on the sensitivity of the observations and the distance of the galaxy. However, a fair comparison between different studies requires the application of the incompleteness corrections we described in this work.


\subsection{Multi - wavelength comparison}
\par By construction of our sample, the four galaxies we consider in this study have been observed in the X-rays with the \textit{Chandra} X-ray  Observatory. These observations reach limiting luminosities of  $3\times 10^{37}\,\rm  erg\, s^{-1}$, $6.8\times 10^{36}\, \rm  erg\, s^{-1}$ ,  $7.4\times 10^{36}\, \rm  erg\, s^{-1}$, and $3.3\times 10^{36}\, \rm  erg\, s^{-1}$  for  NGC 45, NGC 55, NGC 1313, and NGC 7793 respectively (\citealt{2016ApJ...829...20W}). As we would expect the fainter limit is higher for the more distant galaxies. Next we discuss the multi-wavelength (X-ray, radio, optical) populations of SNRs in our sample galaxies, focusing on published samples of SNRs in the X-ray and radio bands. A more detailed study of the X-ray emission of SNRs based on a reanalysis of the existing data will be published in a forthcoming paper. 
In \autoref{table:comparison_others} we present the candidate SNRs of the galaxies of our study, along with their counterparts from other studies. Following, we discuss in more detail these comparisons.
In NGC 45, we find that none of the optically detected SNRs has any counterpart in radio or X-ray wavelengths (\citealt{2015AJ....150...91P}). In NGC 55, 
only source 1 from the possible candidate SNRs has also emission in X-rays (source 131 in \citealt{2015AJ....150...94B}, or source 85 in \citealt{2006MNRAS.370...25S}). The fact that in the optical band we detect only 1 source out of the 18 X-ray SNRs, is probably an age effect {\citep[although we cannot exclude the possibility of ISM differences, especially given the fact that NGC\,55 has different morphology from the other 3 galaxies in our sample; c.f.   ][]{2010ApJ...725..842L}}. The blast waves of very young SNRs heat the material behind the shock front to very high temperatures ($\rm \sim 10^{7}\, K$) producing thermal X-rays (e.g. \citealt{2020ApJ...901..119R,2016A&A...585A.162M,2010ApJ...725..842L}). Most SNRs that produce X-ray emission are in the phase of free expansion or in the early adiabatic phase (e.g. \citealt{2012A&ARv..20...49V}) where we do not expect strong optical emission. Moreover, the SNR J001514-391246 in \citet{2013Ap&SS.347..159O}, which emits in radio and X-rays, presents a shell-like structure ($\rm \sim 44\,pc$) in our H$\rm \alpha$ image. Such SNRs would not be detected efficiently in our survey, which focuses on point-like sources. However, we find that this specific SNR does not show strong [\ion{S}{II}] emission indicating that it could be embedded in \ion{H}{II} region.

\par  In NGC 1313, we do not detect any of the young X-ray emitting SNRs  reported in \citet{1995ApJ...446..177C}. We see strong H$\rm \alpha$ emission at the position of SN 1978K which emits in the X-rays (\citealt{smith2007,1995ApJ...446..177C,1996ApJ...456..187S,1994PASJ...46L.115P})
and in radio (\citealt{1994A&A...285..687A}), but its low [\ion{S}{II}] emission  indicates that the shock may have not excited enough material to make it visible as an optical SNR.
\par In NGC 7793 the candidate SNR 22 coincides with a radio SNR of the work of \citet{2002ApJ...565..966P}. Two of the 3 optically selected radio SNRs of \citet{2014ApSS.353..603G} (29, 31) coincide with our candidate SNRs 22 and 19 respectively. The third optically selected radio SNR of the same work (58) has been also identified in our study but with a significance below 2$\rm \sigma$ so it is not considered here. The larger structure 6 has been detected in the radio (\citealt{2002ApJ...565..966P}) and in X-rays (\citealt{2011AJ....142...20P}). However, we do not consider this structure as an SNR but rather as a super-bubble because of its large size ($\sim$ 320 pc). This is confirmed by the work of \citeauthor{2012MNRAS.427..956D} (\citeyear{2012MNRAS.427..956D}) who suggest that it is a super-bubble created by a microquasar. 


\begin{table}
\caption{Comparison with other studies}

\begin{tabular}{cccc}
ID & RA & Dec  & Other surveys\\
& (J2000) & (J2000) &    \\
 & hh:mm:ss & dd:mm:ss &  \\
\hline
\multicolumn{4}{c}{NGC 55}\\		
\hline
1 & 00:15:43.2 & -39:16:03.0 & 131$^*$ (\citealt{2015AJ....150...94B}) \\
\hline
\multicolumn{4}{c}{NGC 7793}\\
\hline
1  & 23:57:45.0 & -32:37:40.2 & S8  (\citealt{Blair1997})\\
3  & 23:58:06.5 & -32:35:37.0 & S28 (\citealt{Blair1997})\\
4  & 23:57:38.7 & -32:34:38.5 & S1  (\citealt{Blair1997})\\
6  & 23:57:52.6 & -32:33:54.4 & S14 (\citealt{Blair1997})\\
9  & 23:57:41.1 & -32:37:02.1 & S5  (\citealt{Blair1997})\\
13 & 23:57:54.5 & -32:35:12.2 & S16 (\citealt{Blair1997})\\
18 & 23:57:56.1 & -32:37:18.5 & S20 (\citealt{Blair1997})\\
19 & 23:57:48.3 & -32:36:55.2 & S12 (\citealt{Blair1997})\\ & & & 31$^{**}$ (\citealt{2014ApSS.353..603G})\\
20 & 23:57:51.2 & -32:36:31.7 & S13 (\citealt{Blair1997})\\
21 & 23:57:59.2 & -32:36:06.0 & S24 (\citealt{Blair1997})\\
22 & 23:57:47.3 & -32:35:23.9 & S11 (\citealt{Blair1997})\\ & & &  29$^{**}$ (\citealt{2014ApSS.353..603G})\\ & & &  S11 (\citealt{2002ApJ...565..966P})\\
23 & 23:57:45.8 & -32:35:01.7 & S10 (\citealt{Blair1997})\\
\hline

\end{tabular}
\label{table:comparison_others}
{\hspace{1cm} The sources from the work of \citet{Blair1997} are optical SNRs.\\
$^*$ X-ray SNRs\\
$^{**}$ Radio SNRs}
\end{table}



\subsection{The H$\rm \alpha$ Luminosity Function of SNRs}
\begin{figure}
    \includegraphics[width=0.5\textwidth]{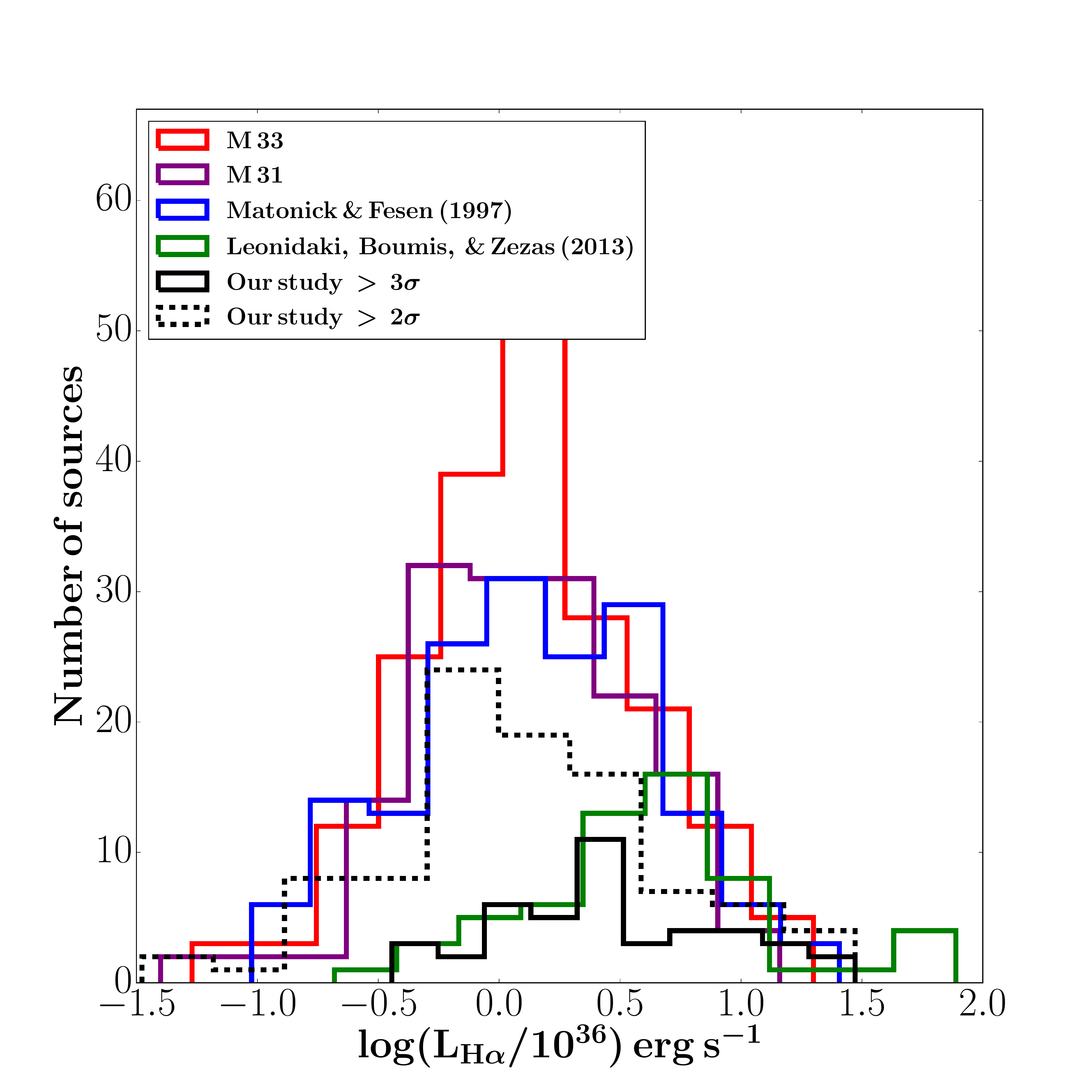}
    \caption{\label{fig:LHa_comp_surv}{{Comparison of the ${\rm{H\alpha}}$ luminosity distribution of our candidate SNRs (black solid line), our possible candidate SNRs (black dashed line) and the studies of spectroscopic SNRs from \citealt{2013MNRAS.429..189L} (green line), and photometric SNRs from \citealt{1997ApJS..112...49M} (blue line), M31 (\citealt{2014ApJ...786..130L}; purple line), and M33 (\citealt{2014ApJ...793..134L}; red line).}}}
\end{figure}

In \autoref{fig:LHa_comp_surv},  we compare the ${\rm{H\alpha}}$ luminosity  distribution of our candidate SNRs (black solid line) and the possible candidate SNRs (black dashed line), with the spectroscopic SNR population in the study of \citeauthor{2013MNRAS.429..189L} (\citeyear{2013MNRAS.429..189L}; NGC 2403, NGC 4212, NGC 3077, NGC 4395, NGC 4449, NGC 5204), and  photometric SNRs in the studies of \citeauthor{1997ApJS..112...49M} (\citeyear{1997ApJS..112...49M}; NGC 5585, NGC 6946, M81, M101), \citeauthor{2014ApJ...786..130L} (\citeyear{2014ApJ...786..130L}; M31) and \citeauthor{2014ApJ...793..134L} (\citeyear{2014ApJ...793..134L}; M33). 

Our candidate SNRs, which is the more secure sample of the two, seem to more closely follow  the distribution of the spectroscopic SNRs of \citeauthor{2013MNRAS.429..189L} (\citeyear{2013MNRAS.429..189L})  while the distribution of the possible candidate SNRs agree with the distribution of the photometric SNRs of the other studies. This is expected since our sample of candidate SNRs consists of brighter objects (since they have higher signal to noise ratio in the [\ion{S}{II}]/H$\rm \alpha$ ratio), and by necessity, spectroscopic surveys also target brighter objects.

In general, the distribution of the ${\rm{H\alpha}}$ luminosity of our candidate SNRs appear to  be  flatter compared to the other studies. This is the result of the strict selection criterion, [\ion{S}{II}]/H$\rm \alpha$ ratio to be 3$\rm \sigma$ above the 0.4 threshold, which is not used in the other surveys. Indeed, we see that the less accurate sample ([\ion{S}{II}]/H$\rm \alpha$ ratio  2$\rm \sigma$ above the 0.4 threshold; black dashed line), agrees more with the other studies.


\par What distinguishes our work from the previous efforts is that: (a) we account for the effects of incompleteness, which allows us to study more accurately the faint end of their distribution and (b) we provide a quantitative description of the LFs, which can be used for comparison between different samples and theoretical models for the population of SNRs in different environments.

\par In \autoref{fig:lf_Ha-both}, we see the incompleteness-corrected luminosity distribution (gray histogram) and the best fit function (blue line) along with the initial, uncorrected data (black-line histogram). We see that apart from the different amplitudes, the mean value of a function that would describe the incompleteness uncorrected data would be shifted to the higher luminosities (since the incompleteness correction affects less the brighter sources) and the skewness would be different. 

\par The skewness at the faint end of the LFs is due to incompleteness. In principle, the limited sensitivity of our survey and the selection of SNRs based on the [\ion{S}{II}]/H$\rm \alpha$ criterion may prevent us from observing the peak of the H$\rm \alpha$ LF of the overall SNR population, even for the nearest galaxies. In order to compare our results with the published photometric SNR samples, which do not impose any threshold on the significance of the [\ion{S}{II}]/H$\rm \alpha$ ratio, we use our less secure sample (> 2$\rm \sigma$), and we compare it with the luminosity distribution of SNRs in the Local Group M31 and M33 galaxies. We see  (\autoref{fig:LHa_comp_surv}) that both show a peak  at $\rm L_{H\alpha} \sim 10^{36}\, erg\,s^{-1}$.

Taking into account the fact that the two surveys have similar flux sensitivity limits, while our objects lie at much larger distances, this strongly suggests that the peak in the H$\rm \alpha$ LF  of SNRs,  selected on the basis of the [\ion{S}{II}]/H$\rm \alpha$ criterion (at signal to noise ratio > 2 $\rm \sigma$ resulting in small contamination by \ion{H}{II} regions), is $\rm \sim 10^{36}\, erg\,s^{-1}$ as measured from our quantitative analysis. Focusing on results, e.g. by integrating the LF, we find that we are missing $\sim$ 30\% of the SNR population down to our detection limit.


\par \citet{Vucetic2015} followed a similar approach for calculating the overall population of SNRs, by integrating a Gaussian distribution fitted to the bright-end of the H$\rm \alpha$  luminosity distribution of the photometric SNR populations in nearby galaxies. However this approach does not account for the effects of incompleteness which may impact even the higher luminosity objects.
 
\par In spiral galaxies  $\sim 75\%$ of SN is core collapse  (\citealt{vink2020}).  We expect that the SNRs that we find in star-forming regions (such as those we consider here) are core collapse SNRs. Hence, since the majority of these SNRs depict the end-point of massive-stars life, they can be used as star formation rate (SFR) indicators. However, because of selection effects and incompleteness biases  it cannot be used directly to derive the relation between SNRs and SFR, unless we correct for the incompleteness. The approach we have used in this work, can be used also to calculate a more reliable relation between the number of SNRs and SFR.

\subsection{The bivariate LF and the degree of excitation}
Observationally, because of the weakness of the [\ion{S}{II}]$\rm \lambda \lambda 6717,6731\AA$ lines, the [\ion{S}{II}]/H$\rm{\alpha}$ criterion generally poses a higher threshold in the SNR luminosity than the sensitivity of the H$\rm \alpha$ data. In fact, the [\ion{S}{II}]/H$\rm{\alpha}$ selection could be the main source of the skewness observed in the  H$\rm \alpha$ LF (\autoref{fig:lf_Ha-both}). Similarly, this selection also results in the skewness observed in the ([\ion{S}{II}]-H$\rm{\alpha}$) bivariate LF along the main axis ([\ion{S}{II}]/H$\rm{\alpha}$; \autoref{fig:LF_3d_all}). The truncation observed along the perpendicular (excitation) axis (\autoref{fig:distances_dist}) is the result of the sharp [\ion{S}{II}]/H$\rm \alpha$ > 0.4 threshold.

\par Regarding the excitation function, as we see in \autoref{fig:lines}, there is a trend for sources with higher H$\rm \alpha$ luminosities to have increasingly lower [\ion{S}{II}]/H$\rm{\alpha}$ ratios (i.e. lower excitation) as is indicated by the sub-linear ($\alpha = -0.09$) slope of the $\rm log(L_{[S\,II]}) - log(L_{H\alpha})$ relation in \autoref{fig:Ha_dist}. This effect would indicate that more luminous objects have lower excitation, which is counter-intuitive since these are expected to be the younger SNRs, and hence the ones with faster shocks and higher excitation. On the other hand, it is more likely that this is an observational effect: luminous SNRs are more likely to be in regions with high ISM density and possibly embedded in \ion{H}{II} regions (e.g. \citealt{2005MNRAS.360...76A}). Indeed, this is a trend that we see in \autoref{fig:incompleteness}, where more luminous SNRs are located in regions with higher backgrounds. We note that this cannot be a selection bias, since brighter objects should be detectable in lower backgrounds as well. This is also one of the reasons (along with the incompleteness due to the [\ion{S}{II}] and the strict selection criterion for the [\ion{S}{II}]/H$\rm \alpha$ at 3$\rm \sigma$), that the completeness is not 100\% even for the brighter sources, while is low at the faint end of the H$\rm \alpha$ LF (\autoref{fig:lf_Ha-both}), where the sources are found in lower background regions. The contamination by the underlying emission from the \ion{H}{II} regions would result in higher H$\rm \alpha$ luminosities but disproportionally lower [\ion{S}{II}] luminosities due to the lower [\ion{S}{II}]/H$\rm{\alpha}$ ratios of the \ion{H}{II} regions. 

\par At first glance this may sound at odds with our expectations that more luminous SNRs are expected to energise more efficiently their surrounding medium. However, this could also be a result of the [\ion{S}{II}]/H$\rm{\alpha}$ selection criterion which biases against lower excitation objects. Only including a more complete population of objects (e.g. based also on the [\ion{O}{I}]/H$\rm \alpha$ criterion e.g. \citealt{2020MNRAS.491..889K}; \citealt{Fesen1985}) we may draw more robust conclusions on the excitation function and the degree of energization of the local ISM.

\par
\begin{figure}
    \includegraphics[width=0.5\textwidth]{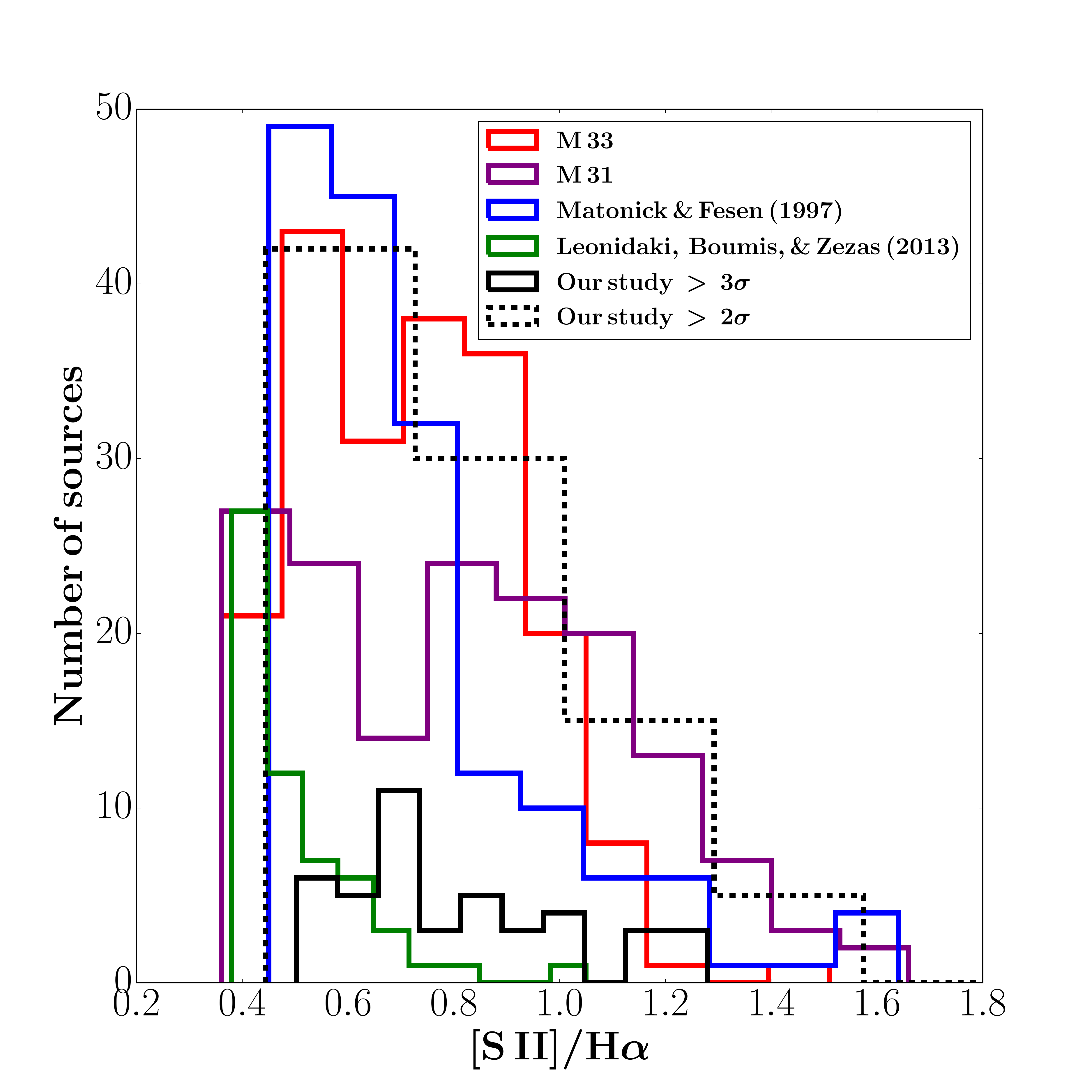}%
    \caption{\label{fig:ratio_comp_surv}{{Histograms of the distribution of the [\ion{S}{II}]/H$\rm{\alpha}$ ratios for the SNRs in our sample (black solid line for the candidate SNRs; black dashed line for the total sample, i.e. the candidate and  the possible candidate SNRs),   \citeauthor{2013MNRAS.429..189L} (\citeyear{2013MNRAS.429..189L}; green line), \citeauthor{1997ApJS..112...49M} (\citeyear{1997ApJS..112...49M}; blue line), M31 (\citealt{2014ApJ...786..130L}; purple line), and M33 (\citealt{2014ApJ...793..134L}; red line).}}}
\end{figure}

 In \autoref{fig:ratio_comp_surv}, we compare the distribution of the [\ion{S}{II}]/H$\rm{\alpha}$ ratios of the candidate SNRs (black solid line) and the possible candidate SNRs (black dashed line) detected in our survey and those reported in the surveys  of \citet{2013MNRAS.429..189L}, \citet{1997ApJS..112...49M}, and \citet{2014ApJ...786..130L, 2014ApJ...793..134L}, aiming to examine the SNR excitation between different galaxies. The range of the  [\ion{S}{II}]/H$\rm{\alpha}$  values of our possible candidate SNRs is more similar to the values of the SNRs of the other surveys, although the distributions are not the same, while the [\ion{S}{II}]/H$\rm{\alpha}$ ratios of our candidate SNRs is quite different. 

\par The difference, between our  sample and the other studies in the faint end has to do with the different adopted [\ion{S}{II}]/H$\rm{\alpha}$ ratio to identify  SNRs (for our possible candidate SNRs), and with the fact that our selection criterion in the [\ion{S}{II}]/H$\rm{\alpha}$ ratio (3$\rm \sigma$ above 0.4) is satisfied only for sources with high [\ion{S}{II}]/H$\rm{\alpha}$ ratios (for our candidate SNRs). The general difference in the [\ion{S}{II}]/H$\rm{\alpha}$ ratio distributions, between our possible candidate SNRs and the other surveys (that do not impose any threshold at the significance level of the [\ion{S}{II}]/H$\rm{\alpha}$ ratio), may have to do with the different ISM conditions of the galaxies, but also with the different sensitivity between the studies. This emphasises the need for incompleteness corrections also in the [\ion{S}{II}]/H$\rm \alpha$ ratio distributions.

\subsection{The effect of [S\, \small{II}\large{]}/H$\rm{\alpha}$ > 0.4 criterion} \label{effect}
Correcting our data for incompleteness, gives us the possibility to recover a number of SNRs that we did not detect (due to their faintness, their galactic environment etc.) down to our faintest detection limit. However, there is a number of SNRs that we do not detect due to selection criteria: for example, the [\ion{S}{II}]/H$\rm{\alpha}$ > 0.4 criterion which is biased against low velocity SNRs.

\par If the SNRs with [\ion{S}{II}]/H$\rm{\alpha}$ < 0.4 follow the extension of the distribution that we derived from those  with [\ion{S}{II}]/H$\rm{\alpha}$ > 0.4, we can estimate the fraction of SNRs that we miss because of the [\ion{S}{II}]/H$\rm{\alpha}$ selection criterion. Then, the missing population is given by the extrapolation of the best-fit truncated Gaussian (\autoref{fig:distances_dist}) to lower [\ion{S}{II}]/H$\rm{\alpha}$ ratios, after removing the truncation term. Applying this method on the best fit truncated Gaussian to the excitation function of SNRs, we find that almost 47\% of the overall SNR population is not accounted for, under the assumption that the distribution of [\ion{S}{II}]/H$\rm{\alpha}$ > 0.4 extends to lower [\ion{S}{II}]/H$\rm{\alpha}$ flux ratios.

\par However, the [\ion{S}{II}]/H$\rm{\alpha}$ > 0.4 criterion is an empirical diagnostic that comes from studies of SNRs in the SMC and the LMC (\citealt{Mathewson&Clarke}).  As discussed in \citet{2020MNRAS.491..889K}, such a criterion may miss legitimate SNRs with relatively low shock velocities which result in lower [\ion{S}{II}]/H$\rm{\alpha}$ ratios (\citealt{2008ApJS..178...20A}). Such low velocities are characteristic of older SNRs (e.g. \citealt{2011piim.book.....D}) or they may reflect their local ISM conditions (e.g. SNRs expanding in dense environments lose quickly their momentum and they slow down faster; e.g. \citealt{2019MNRAS.488..978J};  \citealt{1988ApJ...334..252C}).  Therefore, in order to have a more reliable picture of the overall SNR population one would need to combine fast-shock sensitive indicator (such as [\ion{S}{II}]/H$\rm{\alpha}$) with a slow-shock indicator (such as the [\ion{O}{I}]/H$\rm{\alpha}$; \citealt{2020MNRAS.491..889K}; \citealt{Fesen1985}).

\section{Conclusions} \label{conclusion}
In this work, we present the systematic study of optical SNR populations in 4 nearby spiral galaxies: NGC 45, NGC 55, NGC 1313, and NGC 7793. We identify 42 candidate and 54 possible candidate SNRs (84 of which are new identifications) detected at the 3$\rm \sigma$ level in deep H$\alpha$ images and selected on the basis of their [\ion{S}{II}]/H$\alpha$ flux ratios (higher than 0.4 at the 3$\rm \sigma$ and 2$\rm \sigma$ level respectively) following a fully automated procedure. 

\par Based on this sample, we calculate the H$\alpha$ luminosity function and the joint H$\alpha$-[\ion{S}{II}] luminosity function after accounting for incompleteness effects. The latter presents a new way to encode the luminosity and excitation of the SNR populations. We model the luminosity function with a skewed Gaussian distribution. We find that the H$\alpha$ luminosity at the peak (mean) of the distribution of the overall sample (candidate and possible candidate SNRs) is $\rm \sim 10^{36}\, erg\,s^{-1}$. This is consistent with the peak of the H$\alpha$ luminosity distribution in SNRs detected in the M31 and M33 spiral galaxies for the same quality of data (also less accurate sample).

This is a pilot study to demonstrate the automated method for the identification of  SNRs, the two new metrics of their SNR populations we introduced, and the application of  incompleteness effects. This method will be applied to a larger and more diverse sample of galaxies in order to compare the characteristics of SNR populations in diverse environments.

We also find that SNRs with higher H$\alpha$ luminosity, tend to have lower excitation (i.e. lower [\ion{S}{II}]/H$\alpha$ ratio), although this could be an environmental effect. This  can be seen directly by our data, where brighter sources are located in regions with higher backgrounds.  The excitation function of the overall SNR population is modeled by a truncated Gaussian with the truncation accounting for the [\ion{S}{II}]/H$\alpha$ selection effect. We find that the width of the Gaussian is 0.11 in terms of distance (from the line $\rm y = -0.09x + 0.26$).


\section*{Acknowledgements}
We thank the anonymous referees for helpful comments that helped
to improve the clarity of the paper. We acknowledge funding from the European Research Council under the European Union's Seventh Framework Programme (FP/2007-2013)/ERC Grant Agreement n. 617001. This project has received funding from the European Union's Horizon 2020 research and innovation programme  under the Marie Sklodowska-Curie RISE action,  grant agreement No 691164 (ASTROSTAT). We also acknowledge support from the European Research Council under the European Union’s Horizon 2020 research and innovation program, under grant agreement No 771282. The observations made at Cerro Tololo Inter-American Observatory at NSF’s NOIRLab (NOIRLab Prop. ID: 2011B-0550; PI: A. Zezas), which is managed by the Association of Universities for Research in Astronomy (AURA) under a cooperative agreement with the National Science Foundation. IL acknowledges support by Greece and the European Union (European Social Fund - ESF) through the Operational Programme "Human Resources Development, Education and Lifelong Learning" in the context of the project "Reinforcement of Postodoctoral Researchers-2nd Cycle" (MIS-5033021), implemented by the State Scholarships Foundation (IKY). We also want to thank John Raymond for the useful discussion about shock models and physical properties of SNRs.

\section*{DATA AVAILABILITY}

The data underlying this article are available in the article and in its online supplementary material.
The raw data are available from the NOAO  data archives.




\twocolumn



\newpage
\onecolumn
\appendix
\section{Possible candidate SNRs (<$\rm 3\sigma$)} \label{SNR_2s}
In \autoref{table:SNRs_total_2s} we present the possible candidate SNRs, i.e. SNRs with 2$\rm \sigma$ <  [\ion{S}{II}]/H$\rm \alpha$ - 0.4 < 3$\rm \sigma$. The first column gives the ID of the source, the second and the third  columns give the RA and Dec coordinates (in J2000), the fourth and fifth columns give the H$\rm{\alpha}$ and [\ion{S}{II}] fluxes with their uncertainties, and the sixth column gives the [\ion{S}{II}]/H$\rm{\alpha}$ ratio with its uncertainty.
\begin{ThreePartTable}
\begin{longtable}{cccccc}
\caption{Possible candidate SNRs}\\
\hline 
ID & RA & Dec & $\rm{F_{H\alpha}} \pm \rm{\delta F_{H\alpha}}$ & $\rm{F_{[\ion{S}{II}]}} \pm \rm{\delta F_{[\ion{S}{II}]}}$ & $\frac{\rm{F_{[\ion{S}{II}]}}}{\rm{F_{H\alpha}}} \pm $ ($\rm{\delta}\frac{\rm{F_{[\ion{S}{II}]}}}{\rm{F_{H\alpha}}} $)\\ 
 & (J2000) & (J2000) &  &  &  \\
 & hh:mm:ss & dd:mm:ss & ($\rm 10^{-16}\, erg\,s^{-1}\,cm^{-2}$) & ($ \rm 10^{-16}\, erg\,s^{-1}\,cm^{-2}$) &  \\ 
\hline
\multicolumn{6}{c}{NGC 45\,($\rm 2\sigma$)}\\		
\hline
1 & 00:14:02.3 & -23:10:27.5 &    6.3 $\pm$   0.54 &   4.21 $\pm$   0.64 &  0.67 $\pm$  0.10 \\
2 & 00:14:12.4 & -23:10:27.1 &   16.5 $\pm$    0.4 &   7.31 $\pm$   0.30 &  0.44 $\pm$ 0.02 \\
3 & 00:14:09.8 & -23:09:52.4 &   2.79 $\pm$   0.22 &   1.85 $\pm$   0.24 &  0.66 $\pm$  0.10 \\
\hline
\multicolumn{6}{c}{NGC 55\,($\rm 2\sigma$)}\\
\hline
1 & 00:15:43.5 & -39:16:41.3 &   7.48 $\pm$   0.46 &   4.01 $\pm$    0.4 &  0.54 $\pm$ 0.06 \\
2 & 00:15:56.9 & -39:15:45.3 &   1.84 $\pm$    0.6 &   2.76 $\pm$   0.45 &   1.5 $\pm$  0.5 \\
3 & 00:15:29.9 & -39:14:15.0 &    2.8 $\pm$   0.61 &   3.54 $\pm$    0.7 &   1.3 $\pm$  0.4 \\
4 & 00:15:06.9 & -39:12:24.4 &   3.45 $\pm$   0.63 &   3.47 $\pm$   0.64 &     1 $\pm$  0.3 \\
5 & 00:14:54.0 & -39:11:51.5 &   11.4 $\pm$    3.3 &   16.1 $\pm$    2.3 &   1.4 $\pm$  0.5 \\
6 & 00:14:20.2 & -39:10:09.0 &    1.3 $\pm$   0.19 &   1.15 $\pm$   0.22 &  0.88 $\pm$  0.2 \\
7 & 00:14:01.5 & -39:10:21.0 &   3.15 $\pm$   0.24 &   2.05 $\pm$   0.29 &  0.65 $\pm$  0.1 \\
8 & 00:14:40.5 & -39:08:53.6 &   3.13 $\pm$   0.27 &   2.13 $\pm$   0.34 &  0.68 $\pm$  0.1 \\
9 & 00:13:51.4 & -39:08:46.7 &  0.702 $\pm$   0.18 &  0.975 $\pm$    0.2 &   1.4 $\pm$  0.5 \\
\hline
\multicolumn{6}{c}{NGC 1313\,($\rm 2\sigma$)}\\
\hline
1 & 03:18:01.6 & -66:31:30.5 &   0.87 $\pm$   0.24 &   1.77 $\pm$    0.3 &     2 $\pm$  0.7 \\
2 & 03:18:09.3 & -66:31:12.4 &   3.19 $\pm$   0.46 &   2.42 $\pm$   0.35 &  0.76 $\pm$  0.2 \\
3 & 03:18:17.5 & -66:30:20.4 &   5.53 $\pm$   0.98 &   5.58 $\pm$    1.1 &     1 $\pm$  0.3 \\
4 & 03:17:38.0 & -66:30:01.5 &   6.81 $\pm$   0.33 &   3.66 $\pm$   0.32 &  0.54 $\pm$ 0.05 \\
5 & 03:17:47.5 & -66:29:48.5 &   12.9 $\pm$    0.4 &   6.56 $\pm$   0.43 &  0.51 $\pm$ 0.04 \\
6 & 03:18:39.5 & -66:29:12.1 &   13.6 $\pm$    1.2 &   8.47 $\pm$   0.95 &  0.62 $\pm$ 0.09 \\
7 & 03:18:33.1 & -66:28:50.0 &   7.22 $\pm$   0.81 &   5.35 $\pm$   0.66 &  0.74 $\pm$  0.1 \\
8 & 03:18:36.3 & -66:27:58.6 &   4.03 $\pm$   0.43 &   2.97 $\pm$   0.38 &  0.74 $\pm$  0.1 \\
\hline
\multicolumn{6}{c}{NGC 7793\,($\rm 2\sigma$)}\\
\hline
1 & 23:58:08.2 & -32:36:41.2 &     4.3 $\pm$     0.5 &     3.2 $\pm$    0.34 &  0.74 $\pm$  0.1 \\
2 & 23:57:37.2 & -32:36:32.1 &     3.6 $\pm$    0.33 &     2.4 $\pm$    0.36 &  0.67 $\pm$  0.1 \\
3 & 23:57:31.3 & -32:34:50.9 &     1.5 $\pm$    0.25 &     1.8 $\pm$    0.32 &   1.2 $\pm$  0.3 \\
4 & 23:57:54.2 & -32:37:17.3 &     4.1 $\pm$    0.61 &     4.2 $\pm$    0.83 &   1.0 $\pm$  0.2 \\
5 & 23:57:44.6 & -32:37:03.4 &     7.4 $\pm$     0.6 &     4.5 $\pm$    0.62 &  0.61 $\pm$  0.1 \\
6 & 23:57:41.0 & -32:36:54.9 &     5.5 $\pm$    0.52 &     3.7 $\pm$    0.54 &  0.66 $\pm$  0.1 \\
7 & 23:57:38.6 & -32:34:49.1 &     1.1 $\pm$    0.33 &     3.6 $\pm$    0.45 &   3.3 $\pm$  1.0 \\
8 & 23:57:59.3 & -32:34:37.3 &     4.4 $\pm$    0.54 &     3.2 $\pm$    0.59 &  0.74 $\pm$  0.2 \\
9 & 23:57:59.8 & -32:34:26.4 &     7.8 $\pm$    0.87 &     5.5 $\pm$    0.59 &  0.71 $\pm$  0.1 \\
10 & 23:57:37.9 & -32:33:53.2 &     3.1 $\pm$    0.32 &     2.5 $\pm$    0.48 &  0.79 $\pm$  0.2 \\
11 & 23:57:50.6 & -32:33:13.8 &     4.5 $\pm$    0.58 &     3.3 $\pm$    0.52 &  0.73 $\pm$  0.1 \\
12 & 23:57:48.4 & -32:36:42.8 &     3.9 $\pm$    0.79 &     3.8 $\pm$    0.76 &  0.96 $\pm$  0.3 \\
13 & 23:57:58.0 & -32:36:37.7 &     6.1 $\pm$    0.55 &     4.3 $\pm$    0.66 &   0.7 $\pm$  0.1 \\
14 & 23:57:52.2 & -32:36:15.5 &      11.0 $\pm$     0.7 &     7.6 $\pm$     1.1 &  0.69 $\pm$  0.1 \\
15 & 23:57:58.3 & -32:36:08.3 &       9.0 $\pm$     0.7 &     5.5 $\pm$    0.73 &  0.61 $\pm$ 0.09 \\
16 & 23:57:46.0 & -32:36:06.1 &     5.1 $\pm$    0.73 &     6.5 $\pm$     1.2 &   1.3 $\pm$  0.3 \\
17 & 23:57:58.2 & -32:36:02.1 &     2.4 $\pm$    0.42 &     2.5 $\pm$    0.57 &    1.0 $\pm$  0.3 \\
18 & 23:57:58.7 & -32:35:56.4 &     2.2 $\pm$    0.41 &     2.6 $\pm$    0.44 &   1.2 $\pm$  0.3 \\
19 & 23:57:55.0 & -32:35:51.3 &     2.1 $\pm$    0.53 &     3.0 $\pm$    0.61 &   1.5 $\pm$  0.5 \\
20 & 23:57:43.9 & -32:35:33.1 &     2.3 $\pm$    0.65 &     5.5 $\pm$    0.71 &   2.4 $\pm$  0.7 \\
21 & 23:57:44.2 & -32:35:19.4 &     5.5 $\pm$    0.81 &     4.5 $\pm$    0.75 &  0.81 $\pm$  0.2 \\
22 & 23:58:00.4 & -32:35:14.3 &     4.9 $\pm$    0.44 &     3.3 $\pm$    0.45 &  0.67 $\pm$  0.1 \\
23 & 23:57:44.1 & -32:34:55.4 &     5.3 $\pm$    0.86 &     4.7 $\pm$    0.99 &  0.89 $\pm$  0.2 \\
24 & 23:57:50.1 & -32:34:44.4 &       4.0 $\pm$    0.57 &     4.5 $\pm$    0.75 &   1.1 $\pm$  0.2 \\
25 & 23:57:58.4 & -32:34:41.3 &     5.1 $\pm$    0.42 &     3.3 $\pm$     0.5 &  0.64 $\pm$  0.1 \\
26 & 23:57:41.0 & -32:34:35.7 &     2.6 $\pm$    0.38 &     2.5 $\pm$     0.5 &  0.95 $\pm$  0.2 \\
27 & 23:57:51.0 & -32:34:28.9 &     9.2 $\pm$    0.63 &     6.6 $\pm$    0.91 &  0.72 $\pm$  0.1 \\
28 & 23:57:54.2 & -32:34:16.4 &     1.5 $\pm$    0.31 &     1.8 $\pm$    0.39 &   1.2 $\pm$  0.4 \\
29 & 23:57:54.7 & -32:35:35.5 &     2.4 $\pm$    0.66 &     3.3 $\pm$    0.61 &   1.4 $\pm$  0.4 \\
30 & 23:57:58.9 & -32:35:32.2 &     7.4 $\pm$    0.71 &     5.4 $\pm$    0.94 &  0.73 $\pm$  0.1 \\
31 & 23:57:47.2 & -32:34:28.3 &      13.0 $\pm$     1.6 &     8.2 $\pm$    0.97 &  0.62 $\pm$  0.1 \\
\hline
\label{table:SNRs_total_2s}
\end{longtable}

\end{ThreePartTable}


\bsp	
\label{lastpage}
\end{document}